\documentclass[11pt, letterpaper]{article} 

\usepackage[includeheadfoot,
            marginratio={1:1,2:3}, 
            width=440pt, 
            height=688pt,]{geometry}

\usepackage{amsmath}
\usepackage{amsfonts}
\usepackage{amssymb}
\usepackage{bbm,bm}
\usepackage{mathtools}
\usepackage{slashed}
\usepackage{mathalfa}
\usepackage{float}

\usepackage{graphicx,caption,subcaption}
\usepackage{tikz} 
\usepackage{tikz-cd}
\usetikzlibrary{decorations.pathreplacing,calc,tikzmark}
\usepackage{booktabs}
\usepackage{adjustbox}
\usepackage[utf8]{inputenc}
\usepackage[splitrule]{footmisc} 
 
\usepackage{placeins}
\usepackage{empheq}
\usepackage{paralist}
\usepackage{cite}
\usepackage[normalem]{ulem}
\usepackage{soul}

\usepackage[colorlinks=false,urlbordercolor=red
]{hyperref}
\usepackage{xcolor}
\usepackage{tensor}

\newcommand{\nc}{\newcommand}
\nc{\lb}{\llbracket}
\nc{\rb}{\rrbracket}
\nc{\gl}{\llbracket}
\nc{\gr}{\rrbracket}
\nc{\del}{\partial}
\nc{\eq}[1]{\begin{equation}
                     \begin{split} #1 \end{split}
                     \end{equation}
}
\nc{\ov}{\overline}
\nc{\fa}{\hat}
\nc{\fb}{\MakeUppercase}
\nc{\fc}{\tilde}
\nc{\myhash}{\raisebox{\depth}{\#}}

\allowdisplaybreaks[2]
\numberwithin{equation}{section}
\DeclareUnicodeCharacter{2212}{-}

\newenvironment{claim}[1]{\par\textit{Claim:}\space#1}{}

\definecolor{lightyellow}{rgb}{1,1,0.85}
\definecolor{lightblue}{rgb}{0.87,0.94,1}
\definecolor{lightred}{rgb}{1,0.85,0.85}
\definecolor{lightorange}{rgb}{1,0.9,0.8}
\definecolor{lightgreen}{rgb}{0.88,1,0.88}
\definecolor{lightgray}{rgb}{0.97,0.97,0.97}
\definecolor{blue}{RGB}{6,0,255}
\definecolor{green}{RGB}{12, 127, 0}
\definecolor{yellow}{RGB}{211,211,0}
\definecolor{red}{RGB}{253,42,0}
\begin{document}

%%%%%%%%%%% TITLE PAGE %%%%%%%%%%%%%%%

\vspace*{-1.5cm}
\begin{flushright}
  {\small
  MPP-2026-79
  }
\end{flushright}

\vspace{1.0cm}
\begin{center}
  {\huge  Taxonomy of Instanton 
  Corrections in \\[0.2cm] Infinite Distance Limits 
    } 
\vspace{0.4cm}

\end{center}

\vspace{0.25cm}
\begin{center}
{
\Large Manuel Artime$^{1}$, Ralph Blumenhagen$^{1}$, Panagiotis Leivadaros$^{1,2}$

}
\end{center}

\vspace{0.0cm}
\begin{center} 
  \emph{$^{1}$ 
Max-Planck-Institut f\"ur Physik (Werner-Heisenberg-Institut), \\ 
Boltzmannstra\ss e  8,  85748 Garching, Germany } \\[0.1cm] 
\vspace{0.25cm} 
\emph{$^{2}$ Fakult{\"a}t f{\"u}r Physik, Ludwig-Maximilians-Universit{\"a}t M\"unchen, \\ 
  Theresienstr.~37, 80333 M\"unchen, Germany}
\vspace{0.3cm}
\end{center} 
\vspace{0.5cm}

\begin{abstract}
  Using  the BPS-protected higher derivative $R^4$-term as
  an exactly solvable example, we analyze which instanton corrections are
  generated by a one-loop Schwinger integral over the light
  towers of states that arise in infinite distance limits in moduli space. We find that the Schwinger integral fully captures precisely those instantons whose action lies parametrically in the window $(\Lambda_{\rm sp}/M_{\rm light})^{-1} \le {\rm S}_{\rm inst}\le \Lambda_{\rm
    sp}/M_{\rm light}$, that is, instantons whose action is 
  bounded by the ratio of the gravity cutoff and the mass scale
  of the lightest tower. This proposal is supported by considering the 
  entire  moduli space of toroidal compactifications in eight dimensions, together with a number of limits in seven dimensions. In each case, integrating out the light towers via the Schwinger integral reproduces the
  complete contribution of the instantons within the above window. We
  further recast the proposal in terms of the taxonomy
  classification, allowing us to determine the emergent instantonic
  spectrum associated with any infinite distance limit.
  \end{abstract}

\thispagestyle{empty}
\clearpage

\setcounter{tocdepth}{2}
\tableofcontents

%%%%%%%%%%%% INTRO %%%%%%%%%%%%%%%%%%
\section{Introduction}

One of the striking lessons of string theory is that approaching
infinite distance limits in moduli spaces leads to a breakdown of the
low-energy effective field theory (EFT), as at least one infinite
tower of particle-like states becomes exponentially light. This
phenomenon is conjectured to be a universal property of quantum
gravity, formalized by the Swampland Distance Conjecture
\cite{Ooguri:2006in}. It can be thought of as spoiling the validity of
any Wilsonian effective action for this region in moduli space. A refinement of this conjecture is the
Emergent String Conjecture \cite{Lee:2019wij}, which states that the lightest towers
can only be of two types. They either correspond to excitation modes of a weakly coupled critical string, defining an emergent string limit, or to Kaluza–Klein (KK) modes, signaling the decompactification of internal dimensions in an appropriate duality frame.
In both cases, the appearance of these towers lowers  the scale at which quantum gravity (QG) effects
become relevant from the $d$-dimensional Planck mass $M_{\rm
  pl}^{(d)}$ to what is known as the species scale. This scale was introduced in 
\cite{Dvali:2007hz,Dvali:2007wp,Dvali:2010vm} and it is given by
\begin{equation}
    \Lambda_{\rm sp} = \frac{M_{\rm pl}^{(d)}}{N_{\rm sp}^{\frac{1}{d-2}}} \, , 
\end{equation}
where $N_{\rm sp}$ is the number of light species below $\Lambda_{\rm sp}$. In emergent string limits, the species scale corresponds to the mass scale of the associated weakly coupled critical string, whereas in decompactification limits it is set by the higher-dimensional Planck mass.

The species scale also plays a central role in the Emergence Proposal
\cite{Heidenreich:2017sim, Grimm:2018ohb, Heidenreich:2018kpg, Palti:2019pca}, initially
claiming that the dynamics for all fields in the EFT  arise by
integrating out towers of states down from some energy scale below the
Planck scale. First successfully studied in field theoretic approaches
\cite{Marchesano:2022axe,Castellano:2022bvr, Castellano:2023qhp, Blumenhagen:2023yws, Casas:2024ttx}, it was later 
proposed
in \cite{Blumenhagen:2023tev} that the full emergence of the effective
action might be realized in a particular strongly coupled decompactification limit, the
so-called M-theory limit \cite{Blumenhagen:2023xmk}. For an alternative approach see \cite{Hattab:2023moj,Hattab:2024thi,Hattab:2024chf,Hattab:2024ssg}.
The M-theory  claim was tested in
\cite{Blumenhagen:2024ydy} where the entire coupling of the $R^4$-term
in theories with 32 supercharges was obtained by a one-loop Schwinger
integral of those towers of states whose mass scale is not heavier than the
eleven-dimensional Planck mass, the species scale in such a limit.
This analysis of the Emergence Proposal was generalized to  theories
with 16 supercharges \cite{Artime:2025egu}, by considering the  $1/2$-BPS saturated $F^4$-coupling in 
 Type IIA string theory on $K3$, and to theories with
8 supercharges \cite{Blumenhagen:2023tev, Blumenhagen:2025zgf, Artime:2026dmm},  namely Type IIA string theory on Calabi-Yau threefolds. However, a more general understanding of this proposal across other regions of moduli space remains lacking,
which this paper aims to investigate.

One can think of the Emergence Proposal as stating that at asymptotic regions of moduli space
a perturbative QG theory  with a small  parameter $g\ll
1$ arises, leading to
a hierarchical pattern of the mass scales.
In other words, one can distinguish between light and heavy modes, with the
former being the
fundamental quantum degrees of freedom while the latter can be thought of as classical.
In terms of the naturally small parameter $g$, the mass scale of these modes behave as
\begin{equation}
\label{pertandclass}
m_{\rm pert}\simeq g^\alpha \Lambda \,, \qquad\qquad m_{\rm class}\simeq\frac{\Lambda}{g^\beta}
\end{equation}
with $\alpha\ge 0$, $\beta>0$ and $\Lambda$
a characteristic mass scale, which we take to be the species scale $\Lambda_{\rm sp}$.

Note that this is the behavior also known from perturbative quantum
field theories (QFTs), where one also
distinguishes between the classical non-perturbative contributions to
the path integral and the
light quantum fluctuations around them.
Hence, it is in these limits that one finds something like
a perturbative QG theory that shows certain resemblance to quantum
field theory.

From this perspective, the usual perturbative string theory is just
the perturbative QG theory that arises in the
small string coupling regime, $g_s=e^\phi\ll 1$.
Here, the species scale  is given by the string scale $\Lambda_{\rm sp}=M_s$.
The perturbative states are the vibration modes of the fundamental
string and upon compactification
also its KK and winding modes.
On the other hand, the non-perturbative states are the various
$p$-branes with tension
$T_p\simeq M_s^{p+1}/g_s^\beta$, $\beta=1,2$. 
Then, any coupling constant arising in the effective low energy
field theory for the massless modes, has schematically the expansion
\begin{equation}
  \label{pertexpansion}
C_{\rm eff} = \frac{c_0}{g_s^2} +\underbrace{\left(c_1 +{\cal O}\left(e^{-S_{\rm
          ws}}\right) \right)}_{\rm one-loop} +
        {\cal O}(g_s^2)+{\cal O}\left(e^{-S_{\rm st}}\right) \,.
      \end{equation}
That means, there is a classical contribution at string tree level
followed  by, in principle, an infinite series of higher order corrections in
$g_s$. In the corresponding string loop diagrams one sums over the
light towers of perturbative states only. These automatically include
the worldsheet instanton corrections with action $S_{\rm ws}$. In
contrast, space-time instantons come from non-perturbative objects
such as Euclidean $Dp$-branes wrapping internal cycles of the
geometry, with action $S_{\rm st}$, and are not captured
by the loop expansion. Indeed, it is well known that a $Dp$-brane can be described  as a coherent state (boundary state in CFT) of closed strings.

As mentioned, not all infinite distance limits are emergent
string limits and therefore a natural question is whether
a generalization of this picture also holds in more general limits.
If this is  the case, then it will be  reasonable to study which
corrections are already included in the one-loop Schwinger integral
upon summing over 
the light towers of states with a mass scale below or at the species scale.
Analogously, the heavy towers would be considered as
coherent states made out of these perturbative states.

In order to analyze these questions, it is helpful to have a concrete coupling available that is exactly known across the full moduli space. For this reason, we revisit the $1/2$-BPS saturated higher derivative $R^4$-coupling in maximally supersymmetric theories\footnote{These $R^4$-terms have been
  the testing ground for some of the swampland conjectures, such as their relation with the species scale
  \cite{vandeHeisteeg:2023dlw,Castellano:2023aum, Calderon-Infante:2025ldq, Aoufia:2025ppe, Aoufia:2026bau} and to the emergence
  of species scale black hole horizons
  \cite{Calderon-Infante:2023uhz}.}. In fact, this coupling
is known to only receive one-loop and instanton corrections
in string perturbation theory. Moreover, it is believed
to be exactly given by a Schwinger-like one-loop integral
in M-theory, where one sums over {\it all} 1/2 BPS
particle-like states \cite{Green:1997di,Green:1997as}. 

Furthermore, in order to derive lessons that are valid across general
infinite distance limits in the moduli space of these toroidal theories, a
systematic approach is necessary. We make use of the taxonomy
program\footnote{For earlier work see
  e.g. \cite{Calderon-Infante:2020dhm,Etheredge:2022opl,Etheredge:2023odp,
    Calderon-Infante:2023ler, Castellano:2023jjt,Castellano:2023stg}.}
that started in \cite{Etheredge:2024tok} and where the global structure
of infinite distance limits in  moduli spaces was encoded in what was
referred to as tower and species polytopes \cite{Calderon-Infante:2020dhm,Calderon-Infante:2023ler}. Then, for a generic
infinite distance limit one could  straightforwardly identify the species scale and the corresponding light towers from such polytopes. This enables us to  analyze these $R^4$-terms in generic infinite distance limits in the moduli space and extract which instanton corrections are included in the one-loop Schwinger integral\footnote{A group theoretic study of the instantonic corrections has also been put forward in \cite{Green:2011vz}.}. We will find that this follows  the following pattern.
\begin{claim}
    \textit{Integrating out the light towers of states with mass
      scales not heavier than the species scale $\Lambda_{\rm sp}$, \textbf{fully} generates the instanton corrections whose action lies within the bound}
    \begin{equation}\label{window}
        \left(\frac{\Lambda_{\rm sp}}{M_{\rm light}}\right)^{-1}\le {\rm S_{inst}}\le \frac{\Lambda_{\rm sp}}{M_{\rm light}}\,
      \end{equation}
     \textit{with $M_{\rm light}$ denoting the mass scale of the lightest tower.}
   \end{claim}
   
This generalizes the analysis from \cite{Blumenhagen:2024ydy}, where it was found that in the M-theory limit the complete $R^4$-coupling, including the tree-level term from \eqref{pertexpansion} and all the instanton corrections, is encoded in the light towers of particle-like states. We further translate this bound into the language of the taxonomy classification of \cite{Etheredge:2024tok}, supplementing the taxonomy rules with predictive power over what instantonic spectrum arises from the integration over the light towers of states. 

This paper is organized as follows: in Section \ref{section2}, we introduce the one-loop Schwinger integral that generates the $R^4$-coupling under consideration, and illustrate how integrating out states can give rise to instanton corrections. The main part of the text,
Sections \ref{section3} and \ref{section4}, present the main evidence
for our proposal both in eight and seven dimensions respectively. The
reader solely interested in how our proposal can be translated to the
taxonomy language can jump to Section \ref{genericLimits}, where the
proposal is explored in the full eight-dimensional moduli
space. Lastly, in Section \ref{section5} we summarize our findings
and speculate about possible future directions.

\section{Preliminaries}
\label{section2}

To get started, let us review some relevant points about the $R^4$-terms, the relevant Schwinger integrals and how these can include 
non-perturbative instanton corrections.

\subsection{Higher derivative \texorpdfstring{$R^4$}{R⁴}-terms}

In theories with 32 supercharges, the higher derivative $R^4$-term in the effective action is $1/2$-BPS saturated. Consequently, only $1/2$-BPS states contribute to this coupling, while longer supermultiplets, preserving less supersymmetry, do not contribute to it (see e.g.~\cite{deWit:1999ir}). Even though this coupling cannot be computed from a fully fledged quantum theory of M-theory at the moment, it is sufficient to know the $1/2$-BPS spectrum.  In this paper we revisit the coefficient $a_d$ of this higher-curvature term \cite{Green:1997di}
\begin{equation}
    S_{R^4}\simeq M_s^{-1}\mathcal{V}_k\int d^d x\sqrt{-g} \, a_d\,  t_8t_8 \, R^4\,,
\end{equation}
appearing in the $d$-dimensional effective theory arising after compactifying Type IIA string theory on a $k$-dimensional torus $T^k$, with radii $r_i = M_s R_i \,  \text{and volume }  \mathcal{V}_k$ in string units. To keep the presentation simple, we restrict ourselves to rectangular tori and a possible Kalb-Ramond $B_2$-flux is set to zero. While the Schwinger integral is typically written in the M-theory context, here we express it in terms of Type IIA variables. The two descriptions are related by the relations
\eq{
\label{mparameters}
          g_s = (M_* R_{11})^\frac{3}{2}\,,\qquad  M_s^2=M_*^3 R_{11} \,.    
        }

The $1/2$-BPS particle states that can appear in both the eight- and
seven-dimensional setups are bound states of $D0$-branes, wrapped $D2$-branes and wrapped fundamental strings ($F1$) possibly carrying KK
momentum along the compact directions. Let us denote by $m$ the number
of $D0$-branes, $n_{ij}$ the winding number of a $D2$-brane wrapping a $T^2$ along the $(ij)$-plane, $n_i$ the winding number of a fundamental string wrapping the $i$th-direction, and $m_i$ the KK momentum along the $i$th-direction. Then, the masses of this set of particle states are given by
\eq{
  \label{massBPSstates}
  M_{D0}=\frac{M_s}{g_s}\,,\qquad M_{D2_{(ij)}}=\frac{M_s}{g_s} r_i r_j\,,\qquad
   M_{F1_{(i)}}={M_s\, r_i}\,,\qquad M_{{\rm KK}_{(i)}}=\frac{M_s}{r_i}\,.  
}  
The total $1/2$-BPS Schwinger integral for the coefficient of the
$R^4$-coupling in Type IIA units is known \cite{Obers:1998fb} to be given by a Eisenstein series of order $s=k/2-1$ of $E_{k(k)}$, the U-duality group of toroidal compactifications of eleven-dimensional supergravity, which for our purposes reads
\eq{
\label{r47dschwinger}
a_{d}\simeq \frac{2\pi}{\mathcal{V}_k} &\hat{\sum_{m,m_i,n_i,n_{ij}}} \int_0^\infty  \frac{d t}{t^{\frac{4-k}{2}}}\;
\delta({\rm BPS})\, \\
&\qquad\qquad \exp\left(-\frac{\pi t}{M^2_s}\left(
    M^2_{D0} \,m^2 + M^2_{{\rm KK}_{(i)}} m_i^2 + M^2_{F1_{(i)}} n_i^2
  + M^2_{D2_{(ij)}} n_{ij}^2\right)\right)\,,
}
where in the following the hatted-sum indicates that the term with all integers set to zero has been excluded. Furthermore, the states appearing in the Schwinger integral are subject to a set of $1/2$-BPS conditions \cite{Obers:1998fb,Obers:1999um}, which in the Type IIA frame reads
\begin{align}
\label{BPSone}
n_{[i}\,  n_{jk]}&=0\,,\qquad\quad \sum_{j=1}^{3}n_{j}\, m_j=0 \,,\qquad \quad
n_i\, m + \sum_{j=1}^{3}n_{ij}\, m_j  =0\,.
\end{align}

This is a quite involved integral that has UV divergences and
also develops singularities from the infinite sums over
the KK and wrapping numbers. In  \cite{Blumenhagen:2024ydy} a recipe to deal
with these divergences and to implement the Diophantine
$1/2$-BPS constraints was developed. This can be summarized
as a combination of minimal subtraction of UV divergences
of the integral itself and $\zeta$-function regularization
of the diverging infinite sums over integers. For more details
we refer to the original literature. Since the main proposal of this work deals with instanton corrections, let us recall how these arise with an example before proceeding with the complete analysis in eight dimensions.

\subsection{Instanton corrections}

Consider the contribution in seven dimensions to the coupling $a_7$
where one $D2$-brane winding
number $n_{ij}=n$ is non-vanishing and all other integers
except the number of $D0$-branes are also vanishing.
The $1/2$-BPS constraints are all satisfied so that one gets a
contribution
\eq{
a_{7}\simeq \frac{2\pi}{r_1r_2 r_3}
\sum_{n\ne 0} \sum_{m\in\mathbb Z}  \int_0^\infty  \frac{dt}{t^\frac{1}{2}} \;e^{ -\frac{\pi t}{M_s^2}\left(  M_{D0}^2 m^2 + M^2_{D2} n^2 \right)}\,.
}
Performing a Poisson resummation\footnote{\label{footpoissonresum}
For a symmetric positive definite $k\times k$ matrix $G$ and a real
vector $b^I$, Poisson resummation amounts to $$\sum_{m^I\in
  \mathbb{Z}^k} e^{-\frac{\pi}{t}\sum\limits_{I,J}(m^I+b^I)G_{IJ}(m^J+b^J)} =
\frac{t^{\frac k2}}{\sqrt{\det(G_{IJ})}}\sum_{m_I \in\mathbb{Z}^k}
e^{-2\pi i\sum\limits_{I} m_I b^I - \pi t\sum\limits_{I,J} m_I G^{IJ}m_J}.$$} for the sum over $m$, one arrives
at 
\eq{
  a_{7}\simeq  \frac{4\pi}{r_1r_2 r_3}\frac{M_s}{M_{D0}} 
  \sum_{n\ne 0} \sum_{m\in\mathbb  Z}  \int_0^\infty  \frac{dt}{t} \;
   \exp\left( -\frac{\pi}{t} \frac{M_s^2}{M_{D0}^2} m^2  -\pi t  \frac{M^2_{D2}}{M_s^2} n^2 \right)\,.
 }
 Then, for $m\ne 0$ the integral can be evaluated via the general relation 
\begin{equation}
\label{besselrel}
\int_0^\infty \frac{dx}{x^{1-\nu}} \,e^{-{\frac{b}{x}}-cx}=2 \left|  {\frac{b}{c}}\right|^{\frac{\nu}{2}} K_\nu\left(2\sqrt{|b\, c|}\right)\,,
\end{equation}
where $K_\nu(x)$ denotes the modified Bessel-function of order $\nu$.
In this manner, one obtains a contribution to the Schwinger integral
\eq{
  a_{7}\simeq  \frac{16\pi}{r_1 r_2 r_3}\frac{M_s}{M_{D0}} 
  \sum_{n> 0} \sum_{m>0}  K_0\left(2\pi n m \frac{M_{D2}}{M_{D0}}\right) \,.  
}
Of course, this is a simplified computation and the full treatment is
more involved (see \cite{Blumenhagen:2024ydy}), but this already gives the right
picture.
Namely, the modified Bessel functions contain an exponential
of the argument, which in our case is  
\eq{
  2\pi  \frac{M_{D2_{(ij)}}}{M_{D0}}=2\pi r_i r_j =S_{E\!F1_{(ij)}}\,.
}
As indicated, this  is precisely the action of an Euclidean fundamental string instanton
wrapping the 2-cycle $(ij)$ of the $T^3$. In general, by performing the Poisson resummation for different integers appearing in the integral one is able to generate all potential ratios of the mass scales of relatively 1/2-BPS pairs of
towers of states. Considering for instance the ratio
\eq{
  2\pi  \frac{M_{D2_{(12)}}}{M_{D2_{(23)}}}=2\pi \frac{r_1}{r_3} =U_{(13)}\,,
} 
it is clear that in this way one does not only get instanton
contributions but also complex structure dependent contributions.
In the following we treat these contributions on equal footing
and formally include them in the list of instantons.

As mentioned, the Schwinger integral \eqref{r47dschwinger} is believed to give the complete $R^4$-coupling throughout the moduli space of Type IIA toroidal compactifications. Since all $1/2$-BPS states are treated on equal footing this can be thought of as the relevant Schwinger integral in the bulk of moduli space where all masses are parametrically equal and the species scale is close to the $d$-dimensional Planck scale. However, in infinite distance limits, we expect a different picture where one has a distinction between light and heavy towers in which only the light ones are expected to be integrated out in the one-loop Schwinger integral and with the contribution of the heavy towers being redundant, at least for the instanton contributions already captured by summing over the light towers. 

 To support this claim, we proceed as follows for each infinite distance limit. First, we identify the light towers that lie at or below the species scale. Next, we determine which instanton corrections are captured by the one-loop Schwinger integral over these light states. Finally, we verify that the heavy towers indeed yield redundant contributions to such corrections.

In order to identify the light towers we follow two different approaches. On the one hand we make use of the taxonomy rules developed in \cite{Etheredge:2024tok}. This helps identifying the light towers and their associated species scale in generic directions of moduli space from the polytopes provided therein. We review this approach in the next section. On the other hand, we derive the species scale by using the algorithm developed in \cite{Castellano:2021mmx}, which leads to the species scale
\eq{
    \label{speciesscale}
    \Lambda _{\rm sp}\sim M_{\rm Pl}^\frac{d-2}{d-2+k} \prod_{i=1}^k (\Delta m_i)^\frac{1}{d-2+k}
}      
for  $k$ towers of KK modes with mass splittings $\Delta m_i$, $i=1,\ldots,k$ in $d$ dimensions. Of course, both yield the same results. 

\section{Instanton contributions to the \texorpdfstring{$R^4$}{R⁴}-term in 8D}
\label{section3}
We start by testing our claim in eight dimensions, where the moduli space is still simple enough to admit an exhaustive analysis. In the first part of the section we discuss the complete classification of infinite distance limits, while on the second part we make use of this classification to compute the one-loop Schwinger integral over the light states in each infinite distance and confirm our proposal.

 \subsection{Taxonomy of 8D infinite distance limits}
 \label{taxonomySection}
As just alluded to, for  studying general directions in moduli space
we make use of the tools developed in \cite{Etheredge:2024tok}.
Let us briefly review  the relevant aspects  for our work focusing on the eight-dimensional case. 
Note that in the following we always work with canonically normalized moduli fields. Since these results are used solely as a tool for classification, in Appendix \ref{AppendixA} we provide the map between the different conventions.

Central to the framework of \cite{Etheredge:2024tok} are the $\alpha$-vectors, $\vec\alpha_i$, which encode the dependence of tower masses on the moduli and characterize asymptotic directions along which the $i$th-tower of states becomes lightest. They are defined as \cite{Calderon-Infante:2020dhm}
\begin{equation}
\label{alphavectordefinition}
    \vec{\alpha}_i=-\vec\nabla\log\!\bigg(\frac{m_i}{M_{\rm Pl}^{(d)}}\bigg)\,,
\end{equation}
where the gradient is taken with respect to the canonically normalized moduli and $m_i$ denotes the mass of the corresponding tower.

\begin{figure}[ht]
    \centering
    \includegraphics[width=0.5\linewidth]{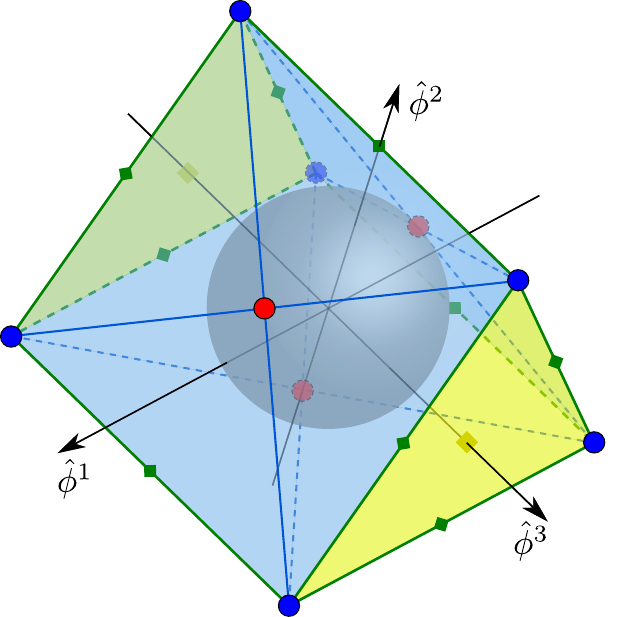}
    \caption{Tower polytope of the maximally supersymmetric eight-dimensional theory \cite{Etheredge:2024tok}. The string towers are depicted in red $\fcolorbox{black}{red}{\phantom{a}}$, while the KK-like towers appear in blue $\fcolorbox{black}{blue}{\phantom{a}}$, green $\fcolorbox{black}{green}{\phantom{a}}$, and yellow $\fcolorbox{black}{yellow}{\phantom{a}}$, depending on whether one, two or three of them become lightest, respectively.}
    \label{polytope3d_8d}
  \end{figure}

These $\alpha$-vectors have been shown to satisfy a series of taxonomy
rules \cite{Etheredge:2024tok,Etheredge:2025ahf}, which allow for a
systematic organization of the different asymptotic directions in
moduli space and lead to the construction of the tower polytope, whose
vertices are given by the $\vec\alpha_i$. The tower polytope encodes
the global structure of infinite distance limits and organizes how the
various asymptotic directions fit together.
For instance, in compactifications to eight dimensions there are three
moduli fields giving rise to the tower polytope  shown in Figure
\ref{polytope3d_8d}.

  Similarly, one can define the species vector \cite{Calderon-Infante:2023ler}
\begin{equation}
    \vec\Lambda_{\rm sp}=-\vec\nabla\log\!\bigg(\frac{\Lambda_{\rm sp}}{M_{\rm Pl}^{(d)}}\bigg)\,,
\end{equation}
which parametrizes the variation of the species scale along moduli
space. The taxonomy rules satisfied by the $\alpha$-vectors lead to
corresponding rules for the species vectors, allowing for the
construction of the species polytope, which up to normalization turns
out to be the dual of the tower polytope.

\begin{figure}[ht]
     \centering
     \begin{subfigure}[b]{0.4\textwidth}
         \centering
         \includegraphics[height=4.5cm]{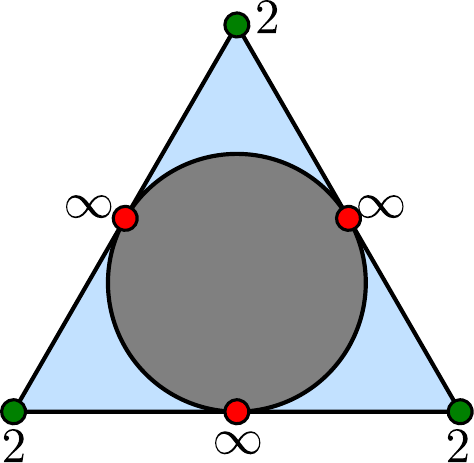}
     \end{subfigure}
     \hspace{10pt}
     \begin{subfigure}[b]{0.4\textwidth}
         \centering
         \includegraphics[height=4.9cm]{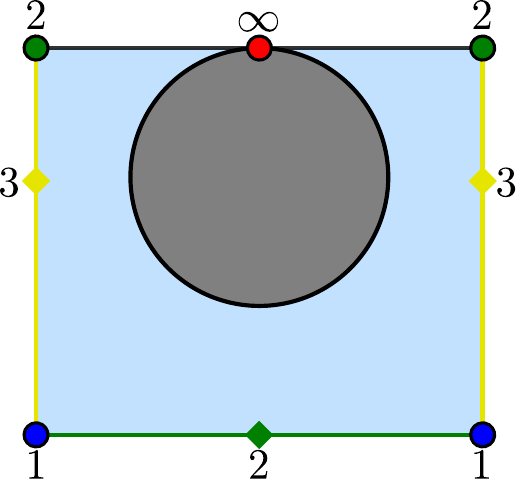}
     \end{subfigure}
        \caption{Two slices of the three dimensional tower polytope given in Figure \ref{polytope3d_8d}.}
        \label{taxonomyTower8d}
\end{figure}

\begin{figure}[ht]
     \centering
     \begin{subfigure}[b]{0.4\textwidth}
         \centering
         \includegraphics[height=3cm]{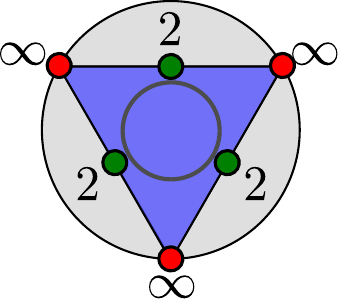}
     \end{subfigure}
     \hspace{10pt}
     \begin{subfigure}[b]{0.4\textwidth}
         \centering
         \includegraphics[height=3.4cm]{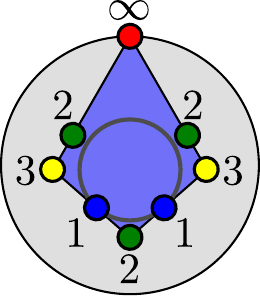}
     \end{subfigure}
        \caption{Two slices of the three dimensional species polytope. Up to normalization it is given by the dual polytope of the slices given in Figure \ref{taxonomyTower8d}.}
        \label{taxonomySpecies8d}
\end{figure}

Figure \ref{taxonomyTower8d} displays two two-dimensional slices of
the three-dimensional tower polytope and Figure
\ref{taxonomySpecies8d} shows the same two slices for the
corresponding species polytope\footnote{We are indebted to the
  authors of \cite{Etheredge:2024tok} for allowing us to use their figures \ref{polytope3d_8d}-\ref{taxonomySpecies8d}.}. The number associated with each vertex indicates how many particle towers become lightest along that direction while vertices labeled by $\infty$ correspond to emergent string limits. The values of the $\alpha$-vectors and $\Lambda_{\rm sp}$-vectors can be found in \cite{Etheredge:2024tok}.

With both polytopes at hand, a general infinite distance limit can be specified by choosing a unit vector $\hat u$. For this direction, the corresponding species vector $\vec\Lambda_{\rm sp}$ is identified with the vector in the species polytope\footnote{In general, the physical species scale does not coincide with the vector on the species polytope, but rather with the largest projection of the species vectors associated to each tower. Since it will not change the analysis in this section, for more details we refer to \cite{Calderon-Infante:2023ler}.}pointing along the same direction $\hat u$, while the projections $\vec\alpha_i\cdot \hat u$ quantify how rapidly the different towers become light along this direction. Comparing these projections to the species scale $|\vec\Lambda_{\rm sp}|$ allows us to determine the hierarchy of light states in a general infinite distance limit.

Let us illustrate how this can be done with a simple example. Consider a direction in moduli space along which the tower of $D0$-branes becomes the lightest tower, corresponding to an M-theory limit. In our conventions, this limit is associated with the bottom-right vertex of the second diagram in Figure~\ref{taxonomyTower8d}. One such direction can be parametrized by the unit vector
\begin{equation}
\hat{u}=\begin{pmatrix}
    -\sqrt\frac{3}{7},-\sqrt\frac{3}{7},-\frac{1}{\sqrt{7}}
\end{pmatrix}\,,
\end{equation}
and its associated species vector is given by $\vec\Lambda_{\rm sp}=\frac{1}{\sqrt{7\cdot 6}}\hat u$. From \cite{Etheredge:2024tok} we can read off the values of $\vec\alpha_i$ for every tower and compute $\vec\alpha_i\cdot \hat u$. In particular, focusing on the towers
\eq{\label{alphavectorsMtheory}
\vec{\alpha}_{D0}&=\begin{pmatrix}
    -\frac{1}{\sqrt{2}},-\frac{1}{\sqrt{2}},-\frac{1}{\sqrt{6}}
\end{pmatrix}\,,\quad \vec{\alpha}_{D2_{(1)}}=\begin{pmatrix}
    0,0,-\frac{1}{\sqrt{6}}
\end{pmatrix}\,,\\
\vec{\alpha}_{{\rm KK}_{(1)}-{\rm KK}_{(2)}}&=\begin{pmatrix}
    \frac{1}{2 \sqrt{2}},-\frac{1}{\sqrt{2}},\frac{1}{2\sqrt{6}}
\end{pmatrix}\,,\quad\quad\! \vec{\alpha}_{D2_{(2)}}=\begin{pmatrix}
    -\frac{1}{2\sqrt{2}},0,\frac{1}{2\sqrt{6}}
\end{pmatrix}\,,\\ 
\vec{\alpha}_{D2_{(12)}}&=\begin{pmatrix}
    -\frac{1}{\sqrt{2}},\frac{1}{\sqrt{2}},-\frac{1}{\sqrt{6}}
\end{pmatrix}\,,
}
we find that
\begin{equation}
    \frac{\vec\alpha_{D0}\cdot \hat u}{|\vec\Lambda_{\rm sp}|}=7\,,\quad \frac{\vec\alpha_{i}\cdot \hat u}{|\vec\Lambda_{\rm sp}|}=1\,,
\end{equation}
where $\vec\alpha_i$ denotes the vectors of the towers considered in
\eqref{alphavectorsMtheory}. As we expected, the tower of $D0$-branes becomes lightest at the fastest rate, while the rest of the towers are at threshold. 

To perform the integrating out procedure it is more convenient to work in the non-canonically normalized scalar fields. In terms of these moduli, infinite distance limits can be described by a parameter $\lambda\to\infty$ with the scalings
\begin{equation}\label{8dscaling}
    g_s\to\lambda^\alpha g_s\,,\quad r_i\to \lambda^{\beta_i}r_i\,.
\end{equation}
The M-theory limit that we have just discussed would then correspond to the values $\alpha=9/7$, $\beta_1=\beta_2=3/7$ up to an overall normalization.

Every infinite distance limit corresponding to a vertex in the slices
of Figures \ref{taxonomyTower8d} and \ref{taxonomySpecies8d} can be
found in terms of these scalings in Table \ref{table_8d}, together
with the corresponding scalings of the masses of both particle and
string towers and their associated species scale. We have named the
limits according to their type of infinite distance limit which will
be elaborated on in more detail in the course of the next subsection.

%%%%%%%%%%%%%%%%%%%%%%%%%
\begin{table}[ht] 
\renewcommand{\arraystretch}{1.2} 
\begin{center} 
\resizebox{\textwidth}{!}{
$\begin{array}{|c ||c|c|c|| c|c|c|c|c|c ||c|c|c|| c|} 
\hline
   & \alpha & \beta_1 & \beta_2  & F1_{(1)} & {\rm KK}_{(1)} &
                                                               F1_{(2)} & {\rm KK}_{(2)} & D0 & D2_{(12)} & F1 & D2_{(1)} & D2_{(2)} & \Lambda_{\rm sp} \\
  \hline \hline
    \text{ESA} & -3 & 0 & 0  & -1 & -1 & -1 & -1 & 2 & 2 & -1 & 1/2 & 1/2 & -1 \\
  \hline
    \text{ESB} & 3/2 & -3/2 & 3/2  & -1 & 2 & 2 & -1 & -1 & -1 & 1/2 & -1 & 1/2 & -1 \\
  \hline
  \text{ESB*} & 3/2 & 3/2 & -3/2  & 2 & -1 & -1 & 2 & -1 & -1 & 1/2 & 1/2 & -1 & -1 \\
  \hline \hline
  
  \text{M-th} & 9/7 & 3/7 & 3/7  & 5/7 & -1/7 & 5/7 & -1/7 & -1 & -1/7 & 2/7 & -1/7 & -1/7 & -1/7 \\
  \hline

  \text{M-th*} & 3/7 & -3/7 & -3/7  & -1/7 & 5/7 & -1/7 & 5/7 & -1/7 & -1 & 2/7 & -1/7 & -1/7 & -1/7 \\
  \hline \hline

  \text{10D} & 0 & 3/4 & 3/4  & 1/2 & -1 & 1/2 & -1 & -1/4 & 5/4 & -1/4 & 1/8 & 1/8 & -1/4 \\
  \hline 

  \text{10D*} & -3/2 & -3/4 & -3/4 & -1 & 1/2 & -1  & 1/2 & 5/4 & -1/4 & -1/4 & 1/8 & 1/8& -1/4 \\
  \hline 

  \text{F-th 1} & 3/2 & 0 & 0 & 1/2 & 1/2 & 1/2 & 1/2 & -1 & -1 & 1/2 & -1/4& -1/4& -1/4\\
  \hline 
  \text{F-th 2} &  -3/4 & 3/4 & -3/4 & 1/2 & -1 & -1 &1/2 &1/2 &1/2 &-1/4 &1/2 &-1/4 &-1/4 \\
  \hline 
  \text{F-th 3} & -3/4  & -3/4 & 3/4 & -1 &1/2 &1/2 &-1 &1/2 &1/2 &-1/4 &-1/4 &1/2 &-1/4 \\
  \hline \hline

  \text{11D} & 1 & 1 & 1 & 1 &-1 &1 &-1 &-1 &1 &0 &0 &0 &-1/3 \\
  \hline 
  \text{11D*} & -1 & -1 &-1  &-1 &1 &-1 &1 &1 &-1 &0 &0 &0 &-1/3 \\
  \hline 
  
\end{array}$}
\caption{Infinite distance limits corresponding to a vertex in the
  two-dimensional slices of Figure \ref{taxonomyTower8d} in terms of
  the scalings given in \eqref{8dscaling}. The scalings of the masses
  of both particle and string states together with the scaling of the
  species scale, computed with \eqref{speciesscale}, are also
  included. The normalization of the scalings has been chosen so that
  the lightest state(s) scales as $\lambda^{-1}$. We have denoted by $^*$ whenever a limit can be obtained by taking a T-duality along both directions of the torus of the non-starred one. }
\label{table_8d}
\end{center} 
\end{table}

\subsection{Taxonomy of instanton corrections}
Let us move on  to the  rather technical part of the paper, where we
integrate out the light towers of states for every infinite distance
limit in Table \ref{table_8d}. For completeness, we first discuss the
case where every particle-like state gets integrated out. This can be
understood as performing the computation in the bulk of moduli space,
where the species scale is of the order of the eight-dimensional
Planck mass and all particle-like towers are at the species scale and
therefore are to be integrated out in the effective description,
resulting in the full coupling in eight dimensions.
Although in the rest we will perform similar computations
for integrating out only parts of the towers of states, readers only interested in the final result can directly
jump to \eqref{8dFull}.

\subsubsection{Instanton expansion in the bulk of moduli space}
The Schwinger integral \eqref{r47dschwinger} over all particle states of the theory is given by
\begin{equation}
\label{allSchwinger}
    a_8 = \frac{2\pi}{t_{12}} \,\hat{\sum_{\substack{ n_1,n_2,n_{12} \\ m,m_1,m_2}}} \int_0^\infty \delta({\rm BPS}) \  \frac{dt}{t} \ e^{-\pi t \left( n_1^2r_1^2+n_2^2r_2^2+ n_{12}^2 \frac{t_{12}^2}{g_s^2} + \frac{m^2}{g_s^2} + \frac{m_1^2}{r_1^2} + \frac{m_2^2}{r_2^2} \right)}\,,
\end{equation}
where $t_{12} = r_1 r_2$ and ${\hat\sum}_{n_1,\ldots,n_k}$  means that
the sum is taken over
all integers $n_i$ with all being zero excluded.

\noindent
Moreover, the BPS condition is given by the system of Diophantine equations
\begin{equation}\label{diophantinesystem8d}
\left\{
\begin{aligned}
    n_1\,m_1 + n_2\,m_2 &= 0 \\
    n_1\,m + n_{12}\,m_2 &= 0 \\
    n_2\,m - n_{12}\,m_1 &= 0
\end{aligned}
\right.\,.
\end{equation}
Following the procedure developed in \cite{Blumenhagen:2024ydy}, we solve this system taking $\{m, m_1, m_2\}$ as unknowns, while keeping $\{n_1, n_2, n_{12}\}$ as fixed coefficients. We further split the contributions in different parametrizations of
\((n_1, n_2, n_{12}, m, m_1, m_2)\), from which we exclude \((0,0,0,0,0,0)\) as some state should be running in the loop.

\begin{enumerate}
    \item \(n_1 = 0 = n_2 = n_{12}\). These states have unrestricted
      KK momentum and no winding along the two directions
      of the torus or wrapped $D2$-branes. 

    \item \(n_1 = n_2 = 0\), \(n_{12} \neq 0\). These states have
      vanishing KK momentum and winding numbers along both
      directions of the torus; however, bound states of $D0-D2$-branes are $1/2$-BPS and thus, can appear.

    \item \(n_1 = 0\), \(n_2, n_{12} \neq 0\) or \(n_2 = 0\), \(n_1, n_{12} \neq 0\). In this case the BPS condition is given by \(n_2\, m + n_{12}\, m_1 = 0\) or \(n_1\, m + n_{12}\, m_2 = 0\). The solutions to these equations are \((0, N \tilde{n}_2, N \tilde{n}_{12}, -M \tilde{n}_{12}, M \tilde{n}_2, 0)\) or \((N \tilde{n}_1, 0, N \tilde{n}_{12}, M \tilde{n}_{12}, 0, -M \tilde{n}_1)\), where $N>0 $ and $ \  M \in \mathbb{Z}$. The tilde notation indicates that the corresponding quantities are coprime, namely $\tilde n_{1,12} = n_{1,12} / \gcd(n_1, n_{12})$.

    \item \(n_{12} = 0\), \(n_1, n_2 \neq 0\). The BPS condition in this case is \(n_1 m_1 + n_2 m_2 = 0\), whose solution is \((N \tilde{n}_1, N \tilde{n}_2, 0, 0, -M \tilde{n}_2, M \tilde{n}_1)\), with \(N > 0\) and \(M \in \mathbb{Z}\).

    \item \(n_{12} = 0\), \(n_1 = 0\), \(n_2 \neq 0\) or \(n_{12} = 0\), \(n_2 = 0\), \(n_1 \neq 0\). These states have one unrestricted KK momentum. The corresponding solutions are \((0, N \tilde{n}_2,0, 0, 0,m_2)\) or \((N \tilde{n}_1, 0,0, 0,  m_2,0)\).

    \item Finally, when all the coefficients \(n_1, n_2, n_{12}\) are non-zero, the most general solution to \eqref{diophantinesystem8d} is \((N \tilde{n}_1, N \tilde{n}_2, N \tilde{n}_{12}, M \tilde{n}_{12}, M \tilde{n}_2, -M \tilde{n}_1)\), with \(N > 0\) and \(M \in \mathbb{Z}\). In this case $\tilde n_{1,2,12}=n_{1,2,12}/\gcd(n_1,n_2,n_{12})$.
\end{enumerate}

We now evaluate the Schwinger integral \eqref{allSchwinger} for these six cases separately and subsequently combine the results. Let us begin with the first case, where \(n_1 = 0 = n_2 = n_{12}\). The integral \eqref{allSchwinger} takes the simple form
\begin{equation}
\label{a0all}
    a^{(0)}_8 = \frac{2 \pi}{t_{12}} \hat{
    \sum_{m,m_1,m_2}} \int_0^\infty \frac{dt}{t}\, e^{-\pi t \left( \frac{m_1^2}{r_1^2} + \frac{m_2^2}{r_2^2} + \frac{m^2}{g_s^2} \right)} \,.
\end{equation}
Next, we decompose the sum as
\(\sum_{m_1,m_2,m\neq(0,0,0)} =
\sum_{\substack{m \neq 0,\, m_1,m_2 \in \mathbb{Z}}}
+
\sum_{m = 0,\, (m_1,m_2) \neq (0,0)}\).
In the first contribution, we perform Poisson resummation over \(m_1\) and \(m_2\), and then isolate the sector with \(m_1 = m_2 = 0\),
\eq{
    a_8^{(0),\,m\neq0}&={2\pi}
    \sum_{m\neq 0} \hat{\sum_{m_1,m_2}}
    \int_0^\infty \frac{dt}{t^2}
    e^{-\frac{\pi}{t}\!\left({m_1^2}{r_1^2}+{m_2^2}{r_2^2}\right)
    -\pi t\frac{m^2}{g_s^2}}
   % \\&+
    +{2\pi}
    \sum_{m\neq 0}
    \int_0^\infty \frac{dt}{t^2}\,
    e^{-\pi t\frac{m^2}{g_s^2}}\,.
}
The first term can be evaluated using~\eqref{besselrel}, whereas the second integral is regularized following the prescription introduced in~\cite{Blumenhagen:2023tev, Blumenhagen:2024ydy}, which amounts to introducing a UV-regulator $\epsilon$, minimal subtraction and regularizing the diverging sums with $\zeta$-function regularization. Concretely, one uses
\begin{equation}
\label{reg2}
\int_{\epsilon}^{\infty} \frac{dt}{t^2}\, e^{-\pi t A}
=
\frac{1}{\epsilon}
+ \pi A \left( \log(\pi A \epsilon) + \gamma_E - 1 \right)
+ \mathcal{O}(\epsilon)\,.
\end{equation}
The contribution corresponding to \(m=0\) can be further decomposed into sectors with \(m_2=0\) and \(m_2\neq 0\),
\begin{equation}\label{all0case}
    a_8^{(0),\,m=0}
    =
    \frac{2\pi}{t_{12}}
    \sum_{m_1\neq 0}
    \int_0^\infty \frac{dt}{t}\,
    e^{-\pi t \frac{m_1^2}{r_1^2}}
    +
    \frac{2\pi}{t_{12}}
    \sum_{m_2\neq 0}  \sum_{m_1\in\mathbb{Z}}
    \int_0^\infty \frac{dt}{t}\,
    e^{-\pi t \left(\frac{m_1^2}{r_1^2} + \frac{m_2^2}{r_2^2} \right)}\,.
\end{equation}
The first integral is again regularized by introducing a UV cutoff,
\begin{equation}
\label{reg1}
\int_{\epsilon}^{\infty} \frac{dt}{t} e^{-tA}
=
-\gamma_E - \log(\epsilon A) + \mathcal{O}(\epsilon)\,.
\end{equation}
With an appropriate choice of regulator\footnote{Following the approach of \cite{Blumenhagen:2024ydy}, the modular invariance of the $R^4$-coefficient in eight dimensions is used to determine the dependence on the regulator. In lower dimensions the regularization procedure is uniquely determined and no ambiguities arise, as the associated Eisenstein series admits a well-defined analytic continuation.},
\(4\pi e^{-\gamma_E}\epsilon\),
the final expression is rendered modular invariant. For the second term in~\eqref{all0case}, we Poisson resum over \(m_1\) and again separate the contributions with \(m_1 = 0\) and \(m_1 \neq 0\). The \(m_1 = 0\) sector is regularized using
\begin{equation}
\label{reg3/2}
\int_{\epsilon}^{\infty} \frac{dt}{t^{3/2}} e^{-tA}
=
\frac{2}{\sqrt{\epsilon}} - 2\sqrt{\pi A}
+ \mathcal{O}(\sqrt{\epsilon})\,,
\end{equation}
while the remaining terms are evaluated with the help of~\eqref{besselrel}. Collecting all contributions, we arrive at
\begin{equation}
\label{allcase0final}
    a^{(0)}_8
    =
    \frac{2 \zeta(3)}{g_s^2}
    - \frac{2 \pi}{t_{12}}
    \log\!\left(
        r_2^2
        \left| \eta(iu) \right|^4
    \right)
    + \frac{8 \pi}{g_s^2}
    \sum_{m>0}\,
    \hat{\sum_{m_1,m_2}}
    \frac{m}{l}
    K_1\!\left(2 \pi m l\right)\,,
\end{equation}
where $u=r_2/r_1$ and
\begin{equation}\label{ED0windings}
    l=\frac{1}{g_s}\sqrt{m_1^2r_1^2+m_2^2r_2^2}\,.
\end{equation}

For the remaining solutions, there exists a convenient method to
perform the calculation simultaneously over the entire set of
states. In particular, we employ the \(\alpha\)-notation\footnote{Let
  $\tilde{n}_\alpha$ denote a $k$-dimensional vector of integers, not
  all necessarily vanishing, indexed by a binary vector $\alpha =
  (\alpha_1,\dots,\alpha_k), \ \alpha_i \in \{0,1\}$. The vector
  $\alpha$ serves as a bookkeeping device that specifies which
  components of the vector $\tilde{n}_\alpha$ are nonzero: the
  condition $\alpha_i = 1$ indicates that $n_i \neq 0$, whereas
  $\alpha_i = 0$ implies $n_i = 0$.} introduced in
\cite{Blumenhagen:2024ydy}. The contribution corresponding to some vector $\alpha$ is given by
\begin{equation}
    a_8^\alpha 
    = \frac{2 \pi }{t_{12}}\cdot 2^{|\alpha|}
    \sum_{\mathbf{\tilde{n}}_\alpha>0} \sum_{N>0} \sum_{M \in \mathbb{Z}}
    \int_0^\infty \frac{dt}{t} 
    \, e^{- \pi t L_\alpha^2 N^2} \, 
    e^{- \pi t M^2 \frac{L_\alpha^2 }{t_{12}^2}} \ ,
\end{equation}
where
\begin{equation}
    L_\alpha^2  = \frac{\tilde{n}_{12}^2}{g_s^2} r_1^2r_2^2 + \tilde{n}_2^2 r_2^2 + \tilde{n}_1^2 r_1^2 \ . 
\end{equation}

We separate the sum into the contributions with \(M = 0\) and \(M \neq 0\). The latter can be evaluated using the identity~\eqref{besselrel}, while the \(M = 0\) term requires regularization according to~\eqref{reg3/2}. Finally, performing the sum over coprime integers using the fact that 
\begin{equation}\label{sumcoprimes}
    \sum_{\mathbf{\tilde{n}_\alpha}>0}1=\left(\frac{1}{2}\right)^{|\alpha|-1}\,,
\end{equation}
we arrive at
\begin{equation}
    a_8^\alpha 
    = (-1)^{|\alpha|} \frac{2 \pi }{t_{12}} 
    \log \!\left( \left|\eta(i t_{12})\right|^4 \right).
\end{equation}
Summing over all admissible configurations, namely over all values of \(\alpha\), one finds
\begin{equation}
\label{allcase1final}
    a_8^{\alpha \neq 0} 
    =\sum_{|\alpha|=1}^3\binom{3}{|\alpha|}a_8^\alpha= - \frac{2 \pi }{t_{12}} 
    \log \!\left( \left|\eta(i t_{12})\right|^4 \right).
\end{equation}
Putting together the contributions in \eqref{allcase0final} and \eqref{allcase1final}, the complete expression for the coefficient \(a_8\) becomes
\begin{equation}
\label{8dFull}
\boxed{a_8
= \frac{2 \zeta(3)}{g_s^2}
   - \frac{2 \pi}{t_{12}}
   \log \!\left(
      r_2^2
      \left| \eta(iu)\, \eta(i t_{12}) \right|^4
   \right) + \frac{8 \pi}{g_s^2}
   \sum_{m>0}
   \hat{\sum_{m_1,m_2}}
   \frac{m}{l}
   K_1 \!\left(
      2 \pi m l
   \right) \,,}
\end{equation}
where \(l\) is defined in \eqref{ED0windings}. From the weakly coupled
Type IIA point of view, this contains a tree-level term in
$g_s$, the one-loop contribution including worldsheet instantons and complex-structure dependent terms\footnote{The exponential dependence for the instantonic terms can be recovered after using the identity
\begin{equation}
    \log\left(|\eta(iu)|^4\right) = - \frac{\pi }{3} u + 4 \sum_{n\geq1} \  \frac{1}{n}  e^{-2 \pi n u }\,.
\end{equation}}, and a non-perturbative piece from $E\!D0$-instantons.

Having available the full $R^4$-term in eight dimensions we can move
forward and, for each of the infinite distance limits in Table $\ref{table_8d}$, test our claim that all the instantons within the bound
\eqref{window} are fully generated by integrating out only the light towers of states.

\subsubsection{M-theory limit/M-theory* limit}

We start by consider the M-theory limit in eight dimensions, i.e a
decompactification limit to nine dimensions. This limit has already
been investigated in this context in
\cite{Blumenhagen:2024ydy,Blumenhagen:2024lmo}, so that we can rely on some of the already obtained results.

The limit is characterized by $D0$-branes being the lightest states,
while KK momentum along both directions of the torus and wrapped
$D2$-branes lie at the species scale, and thus are to be considered
light as well. The $1/2$-BPS condition \eqref{BPSone} reads in this case
\begin{equation}
    n_{12} \, m_2 = 0 \,, \qquad n_{12} \, m_1 = 0 \, .
\end{equation}
Therefore, configurations with KK momentum along a $D2$-brane are not $1/2$-BPS and do not appear in the integrating out procedure. Moreover, as can be seen from Table \ref{table_8d}, the two particle-like states arising from $F1$-strings wrapping internal cycles are heavy in this limit and are therefore considered non-perturbative states. Even though these are not present in the Schwinger integral the light towers of states capture \textit{all} potential instantons, whose action is induced by the ratios of masses of light states
\eq{
S_{U_{(21)}}\sim \frac{m_{{\rm KK}_{(1)}}}{m_{{\rm KK}_{(2)}}}\sim 1\,,\quad S_{E\!D0_{(i)}}\sim \frac{m_{D0}}{m_{{\rm KK}_{(i)}}}\sim \lambda^{-\frac{6}{7}}\,,\quad S_{E\!F1_{(12)}}\sim\frac{m_{D2_{(12)}}}{m_{D0}}\sim\lambda^{\frac{6}{7}}\,.
}
As proposed, these all  lie in the window \eqref{window} with $\Lambda_{\rm sp}/M_{\rm light}\sim\lambda^{6/7}$, so that their total contribution should be given by integrating out the light states. Indeed, this has been explicitly shown in \cite{Blumenhagen:2024ydy} to give the full eight-dimensional $R^4$-coupling \eqref{8dFull}. Thus, integrating out only the light states already yields the full contribution to \(a_8\), and therefore the integration over the heavy states is redundant. For those interested in the details for this computation, we refer to the original literature.

Next, let us move onto the limit denoted by M-th* in Table
\ref{table_8d}. In this limit, the light states are the winding modes
along both directions 1 and 2, together with bound states of
$D0$-branes and wrapped $D2$-branes with the latter being the lightest tower.
This limit can be obtained from the M-theory limit by performing T-duality along both directions of the torus, so we expect to recover again the full coupling. This confirms our claim, since
\begin{equation}
    S_{U_{(12)}} \sim \frac{m_{\mathrm{F1}_{(2)}}}{m_{\mathrm{F1}_{(1)}}} \sim 1\,,\quad S_{E\!F1_{(12)}} \sim \frac{m_{\mathrm{D2}_{(12)}}}{m_{\mathrm{D0}}} \sim \lambda^{-6/7}\,,\quad S_{E\!D0_{(i)}} \sim \frac{m_{\mathrm{D0}}}{m_{\mathrm{F1}_{(i)}}} \sim \lambda^{6/7}\,,
\end{equation}
are within the window \eqref{window}.

Although the computation is quite similar to the M-theory limit, we now show the main steps for illustrative purposes. The Schwinger integral reads
\begin{equation}
    a_8=\frac{2\pi}{t_{12}}\hat{\sum_{n_1,n_2,m,n_{12}}} \int_0^\infty \frac{dt}{t} e^{-\pi t \left(n_1^2r_1^2+n_2^2r_2^2+\frac{m^2}{g_s^2}+n_{12}^2\frac{t_{12}^2}{g_s^2} \right)}\,,
\end{equation}
subject to the $1/2$-BPS conditions
\begin{equation}
    n_1\, m=0\,,\quad n_2 \,m=0\,.
\end{equation}

There are two distinct classes of solutions in this case. The first corresponds to $ m = 0 $ , for which $ n_1 $ and $ n_2$  are arbitrary integers. The second class is given by $ n_1 = n_2 = 0 $, with $m$ being a free integer. For the former, the Schwinger integral reads
\begin{equation}
\label{aMcase1}
a^{(1)}_8
=
\frac{2\pi}{t_{12}}
\hat{\sum_{n_1,n_2,n_{12}}}
\int_0^\infty \frac{dt}{t}\,
e^{-\pi t \left( n_1^2 r_1^2 + n_2^2 r_2^2
+ \frac{r_1^2 r_2^2}{g_s^2} n_{12}^2 \right)} .
\end{equation}
We split the sum as $\hat{\sum}_{n_1,n_2,n_{12}}=\sum_{\substack{n_{12} \neq 0,\, n_1,n_2 \in \mathbb{Z}}}
+
\sum_{n_{12} = 0,\, (n_1,n_2) \neq (0,0)}$. In the first term, we Poisson resum over $n_1$ and $n_2$, and subsequently isolate the contribution corresponding to $n_1 = n_2 = 0$
\eq{
    a_8^{(1),\,n_{12}\neq0}&=\frac{2\pi}{t_{12}^2}\sum_{n_{12}\neq 0} \hat{\sum_{n_1,n_2}}\int_0^\infty \frac{dt}{t^2} \exp\left[-\frac{\pi}{t}\left(\frac{n_1^2}{r_1^2}+\frac{n_2^2}{r_2^2}\right)-\pi t\frac{t_{12}^2}{g_s^2}n_{12}^2\right]
    \\&
    +\frac{2\pi}{t_{12}^2}\sum_{n_{12}\neq 0}\int_0^\infty \frac{dt}{t^2}\, e^{-\pi t\frac{t_{12}^2}{g_s^2}n_{12}^2}\,.
}
The first contribution is computed with the aid of~\eqref{besselrel}, while the second integral is regularized according to the prescription in~\eqref{reg2}. The summand corresponding to $n_{12}=0$ can be split into two other contributions with $n_2=0$ and $n_2\neq 0$
\begin{equation}\label{a8Mtheorydual1}
    a_8^{(1),\,n_{12}=0}=\frac{2\pi}{t_{12}}\sum_{n_1\neq 0}\int_0^\infty \frac{dt}{t}e^{-\pi t n_1^2r_1^2}+\frac{2\pi}{t_{12}}\sum_{n_2\neq 0} \sum_{n_1\in\mathbb{Z}}\int_0^\infty \frac{dt}{t} e^{-\pi t(n_1^2 r_1^2 +n_2^2 r_2^2)}\,.
\end{equation}
We can regularize the former by using~\eqref{reg1}. Importantly, choosing a suitable regulator,  $4\pi e^{-\gamma_E}\,\frac{r_2^2}{r_1^2}\,\epsilon$, allows the final result to be modular invariant. For the second term in \eqref{a8Mtheorydual1}, we perform Poisson resummation over $n_1$ and again split it into the cases $n_1 = 0$ and $n_1 \neq 0$. Furthermore, the  $n_1 = 0$ term is regularized using ~\eqref{reg3/2}, while the remaining terms are evaluated using~\eqref{besselrel}. Putting everything together, we obtain
\begin{equation}
\label{aMcase1final}
    a^{(1)}_8 = \frac{2 \zeta(3)}{g_s^2}
   - \frac{2 \pi}{t_{12}}
   \log \!\left(
      r_2^2
      \left| \eta(iu)\, \right|^4
   \right) 
   + \frac{8 \pi}{g_s^2}
   \sum_{m>0}
   \hat{\sum_{m_1,m_2}}
   \frac{m}{l}
   K_1 \!\left(
      2 \pi m l
   \right) \,,
\end{equation}
with $l$ given by \eqref{ED0windings}.

For the second class of solutions, $n_1=n_2=0$, the Schwinger integral becomes
\begin{equation}
\label{aMcase2}
\alpha^{(2)}_8
=
\frac{2\pi}{t_{12}}
\sum_{n_{12} \neq 0} \sum_{m \in \mathbb{Z}}
\int_0^\infty \frac{dt}{t}\,
e^{-\pi t \left( \frac{m^2}{g_s^2}
+ \frac{r_1^2 r_2^2}{g_s^2} n_{12}^2 \right)} .
\end{equation}
This integral can be evaluated in a similar manner to the other set of solutions. We perform Poisson resummation over $m$ and split the resulting sum into the contributions with $m=0$ and $m\neq0$. The former is regularized using \eqref{reg3/2}, while the latter is evaluated using \eqref{besselrel}. In the end, we obtain
\begin{equation}
\label{aMcase2final}
a^{(2)}_8
=
-\,\frac{2\pi}{t_{12}}
\log\!\left(
\left| \eta(i t_{12}) \right|^4
\right) .
\end{equation}

By combining all these contributions, namely \eqref{aMcase1final} and \eqref{aMcase2final}, we indeed obtain the full $R^4$-coupling in eight dimensions as given in \eqref{8dFull}. One important difference with respect to the computation in the M-theory limit is that, in the present case, the Poisson resummation needs to be performed over the winding modes. In the rest of the limits a similar situation arises, where one needs to choose the appropriate integers to be Poisson resummed  in order to obtain the correct expansion of the coupling.

\subsubsection{Emergent perturbative Type IIA string (ESA)}
The next limit we consider is the standard weakly coupled Type IIA limit, also already studied in \cite{Blumenhagen:2024ydy}, denoted by ESA in Table \ref{table_8d}. The lightest state is given by the fundamental string, while the particle-like light states consist of winding and KK modes along directions 1 and 2. All other particle states are heavy compared to the species scale, which in the emergent string limit is set by the string scale $M_s$. As one could imagine, we expect to obtain the one-loop correction in $g_s$ of \eqref{8dFull}, which includes the complete contribution of worldsheet instantons and complex structure dependence. Indeed, these instantons have action
\begin{equation}
        S_{U_{(12)}} \sim \frac{m_{F1_{(2)}}} {m_{F1_{(1)}}} \sim 1 \ , \quad \, S_{E\!F1_{(12)}} \sim \frac{m_{D2_{(12)}}}{m_{D0}} \sim 1,
\end{equation}
which lies in the window \eqref{window}, since $\Lambda_{\rm
  sp}/M_{\rm light}\sim 1$ for emergent string limits. On the other
hand, the non-perturbative corrections given by
$E\!D0_{(i)}$-instantons
should not be emergent, consistent with the fact that  they lie above the bound \eqref{window}
\begin{equation}
    S_{E\!D0_{(i)}}\sim\frac{m_{D0}}{m_{F1_{(i)}}}\sim\lambda^3\,.
\end{equation}
This can be confirmed by evaluating the Schwinger integral over the light states
\begin{equation}
\label{stringschwingerES1}
    a_8 = \frac{2 \pi}{t_{12}} \hat{\sum_{n_1,n_2,m_1,m_2}} \int_0^\infty
    \frac{dt}{t} \  \delta({\rm BPS}) \ e^{- \pi t \sum_{i=1}^2 \left(\frac{m^2_i}{r_i^2} + n_i^2 r_i^2\right)}
\end{equation}
subject to the BPS condition 
\begin{equation}
    n_1\,m_1+n_2\,m_2 =0\,.
\end{equation}
This computation  has already been done in \cite{Blumenhagen:2024ydy}, leading to
\begin{equation}
\label{ES1}
    a_8 = - \frac{2 \pi}{t_{12}} \log\left( r_2^2 |\eta(iu) \eta(i t_{12})|^4 \right)\,,
\end{equation}
which matches the one-loop contribution in \eqref{8dFull} and captures the worldsheet instantons and the complex structure dependent terms.
Therefore, it is obvious that the contribution of the heavy states is
redundant.

\subsubsection{Emergent perturbative Type IIB string (ESB/ESB*)}

Next, we consider the limits denoted by ESB and ESB* in Table
\ref{table_8d}. Again, the latter can be obtained after applying
T-dualities along both directions of the torus, so in the following we
only analyze the former. In this case, the lightest tower is a
$D2$-brane
wrapping the first direction of the torus, accompanied by fundamental
string winding modes along the same direction, KK modes along the second
direction, and bound states of $D0$- and wrapped $D2$-branes.

The light towers induce the following instanton
corrections\footnote{Following \eqref{BPSone}, the combinations $D2_{(12)}\!-\!{\rm KK}_{(2)}$ and $D0\!-\!F1_{(1)}$ are non-BPS so their mass ratios do not contribute to the induced instanton corrections in this limit.}
\begin{equation}
    S_{E\!D0_{(2)}} \sim \frac{m_{D0}}{m_{\mathrm{KK}_{(2)}}} \sim \frac{m_{D2_{(12)}}}{m_{F1_{(1)}}} \sim 1\,, \qquad
    S_{E\!F1_{(12)}} \sim \frac{m_{F1_{(1)}}}{m_{\mathrm{KK}_{(2)}}} \sim \frac{m_{D2_{(12)}}}{m_{D0}} \sim 1\,,
\end{equation}
and we would be missing the instantons with actions
\begin{equation}
    S_{U_{(12)}} \sim \frac{r_2}{r_1} \sim \lambda^{3}, \qquad
  S_{E\!D0_{(1)}} \sim \frac{r_1}{g_s} \sim \lambda^{-3},
\end{equation}
which do not have actions parametrically equal to $\Lambda_{\rm sp}/M_{\rm light}\sim 1$. 

As it will become apparent in a moment, this emergent string limit (and its dual ESB*) is better understood in the Type IIB frame after applying a T-duality along the first direction (second direction for ESB*), followed by an S-duality. Applying this chain of dualities to the moduli gives the duality map
\begin{equation}
\label{typeIIBmoduli}
g_s' = \frac{r_1}{g_s} , \qquad
M_s' = M_s \sqrt{\frac{r_1}{g_s}} , \qquad r_1' = \frac{1}{\sqrt{g_s r_1}} , \qquad r_2' = r_2 \sqrt{\frac{r_1}{g_s}}\,,
\end{equation}
where primed quantities denote the moduli in the Type IIB frame. In terms of these variables, the scalings given in Table \ref{table_8d} translate into the Type IIB weak-coupling limit
\begin{equation}
g_s'\to\lambda^{-3} g_s' , \qquad r_1'\to r_1' , \qquad r_2'\to r_2'\,,
\end{equation}
with light states and their induced instanton corrections depicted in Table \ref{table_emergeIIB8d}. \begin{table}[ht] 
\renewcommand{\arraystretch}{1.2} 
\begin{center} 
\begin{tabular}{|l|l||l|l|} 
\hline
light towers    & mass scale & instantons & action   \\
\hline \hline
$F1_{(i)}$  &   $M\sim {M_s' r_i'}\sim\lambda ^{-1}$ &  $U_{(12)}$ &
 $S\sim  \frac{m_{{\rm KK}_{(1)}}}{m_{{\rm KK}_{(2)}}}\sim  \frac{m_{F1_{(2)}}}{m_{F1_{(1)}}}\sim 1$ \\
${\rm KK}_{(i)}$  &   $M\sim \frac{M_s'}{r_i'}\sim \lambda^{-1}$ & $E\!F1_{(12)}$
                          & $S\sim \frac{m_{F1_{(1)}}}{m_{{\rm
                            KK}_{(2)}}}\sim \frac{m_{F1_{(2)}}}
                                       {m_{{\rm KK}_{(1)}}}\sim 1$ \\    
  \hline\hline
$F1$  &   $T\sim {M_s'^2 }\sim\lambda
              ^{-2}$ &   &  \\
  \hline
\end{tabular}
\caption{Mass scales of light towers of particles/strings and their
  induced instanton corrections for dual weakly coupled Type IIB string.}
 \label{table_emergeIIB8d}
\end{center} 
\end{table}

This is clearly the weakly coupled emergent string limit of Type IIB string theory, in which we expect to obtain the complete one-loop contribution in $g_s'$, analogously to the perturbative Type IIA infinite-distance limit. Consistent with this expectation, the actions of the relevant euclidean $D$-brane instantons scale as
\begin{equation}
    S_{E\!D(-1)}\sim\frac{1}{g_s'}\sim\lambda^3\,,\qquad S_{E\!D1_{(12)}}\sim\frac{r_1'r_2'}{g_s}\sim \lambda^3\,,
\end{equation}
and are parametrically larger than $\Lambda_{\rm sp}/M_{\rm light}\sim 1$.

Thus, it remains to show that integrating out the light states gives the complete contribution to the worldsheet instantons and $U$-moduli. In the Type IIB frame, dropping the prime notation for simplicity, the Schwinger integral over the light states becomes
\begin{equation}
\label{ES2IIB}
a_8 = \frac{2\pi r_1}{r_2}
\hat{\sum_{n_1,m_1,n_2,m_2}}
\int_0^\infty \delta(\mathrm{BPS})\,\frac{dt}{t}\,
e^{-\pi t \left( n_1^2 r_1^2 + n_2^2 r_2^2
+ \frac{m_1^2}{r_1^2} + \frac{m_2^2}{r_2^2} \right)}\,,
\end{equation}
where the BPS condition takes the form 
\begin{equation}
\label{BPSIIB}
    n_1 m_1 + n_2 m_2 = 0 \,.
\end{equation}
The only difference with the emergent Type IIA string limit is that we now solve the Diophantine equation \eqref{BPSIIB} by treating the integers $\{m_1,n_2\}$ as coefficients and solving for $\{n_1,m_2\}$. Following the same procedure as in the emergent Type IIA string limit, but performing a Poisson resummation over $m_2$, the resulting $R^4$-coupling in this limit is given by 
\begin{equation}
\label{aES2}
    a_8 
    = - \frac{2 \pi r_1}{r_{2}} 
    \log \!\Bigg( 
    r_2^2 
    \left|\eta\!\left(i u \right) 
    \eta(i t_{12})\right|^4 
    \Bigg)\,,
\end{equation}
which is nothing but the complete one-loop result in the Type IIB frame that one expects in an emergent string limit.

\subsubsection{F-theory limits}
The rest of the limits we consider in the following correspond to
decompactification limits. First, let us consider the limit denoted by
F-th 1 in Table \ref{table_8d}. This strongly coupled Type IIA limit
has as particle-like light states wrapped $D2$-branes and
$D0$-branes.
There are also two light strings given by the wrapped $D2$-branes along each direction of the torus, i.e. $D2_{(1)}$ and $D2_{(2)}$. The only induced instanton corrections by these light states are worldsheet instantons with action $S_{E\!F1_{(12)}}\sim m_{D2_{(12)}}/m_{D0}\sim 1$. Note that the complex structure dependent terms are also within the window \eqref{window} since they also scale as $S_{U_{(12)}}\sim r_2/r_1\sim 1$, however, they cannot be generated by the light states that we integrate out. We discuss this issue at the end of the section.

This limit can be brought into a more familiar form by applying T-duality along the first direction, followed by an S-duality, and then T-duality along both the first and second directions of the torus. The duality map between the resulting Type IIB theory and the starting Type IIA theory reads
\begin{equation}
\label{typeIIBmoduliF}
g_s' = \frac{r_1}{ r_2}\,, \qquad
M_s' = M_s \sqrt{\frac{r_1}{g_s}}\,, \qquad
r_1' = \sqrt{r_1 g_s}\,, \qquad
r_2' = \frac{1}{r_2} \sqrt{\frac{g_s}{r_1}}\,.
\end{equation}
Written in terms of these variables, the scalings listed in Table \ref{table_8d} correspond to the following limit in Type IIB
\begin{equation}
g_s' \rightarrow g_s'\,, \qquad
M_s' \rightarrow \lambda^{-1/4} M_s'\,, \qquad
r_1' \rightarrow \lambda^{3/4} r_1'\,, \qquad
r_2' \rightarrow \lambda^{3/4} r_2'\,,
\end{equation}
with light states and corresponding induced instanton corrections displayed in Table \ref{table_emergeF8DIIB}. The coupling does not scale so it is thought to be of order one.

\begin{table}[ht]
\centering
\renewcommand{\arraystretch}{1.2}
\begin{tabular}{|l|l||l|l|}
\hline
light towers & mass scale & instantons & action \\
\hline\hline
${\rm KK}_{(i)} $
& $M \sim \frac{M'_s}{r'_i} \sim \lambda^{-1}$ 
& $U_{(12)}$ 
& $S \sim \frac{m_{{\rm KK}_{(1)}}}{m_{{\rm KK}_{(2)}}} \sim 1$ \\
\hline\hline
$F1$ 
& $T \sim M_s'^2 \sim \lambda^{-\frac{1}{2}}$ 
& 
& \\
$D1$ 
& $T \sim \frac{M_s'^2}{g'_s} \sim \lambda^{-\frac{1}{2}}$ 
& 
& \\[0.1cm]
\hline
\end{tabular}
\caption{Mass scales of light towers of particles and strings, and their induced instanton corrections for the F-theory limit in eight dimensions in the Type IIB frame.}
\label{table_emergeF8DIIB}
\end{table}
It now becomes apparent why we named this limit as F-theory limit, as the fundamental string and the $D1$-string are light strings on equal footing. Moreover, the KK momenta of the torus are signaling that two directions are decompatifying, analogous to what happens in F-theory limits in Calabi-Yau moduli spaces (see e.g. \cite{Corvilain:2018lgw}). 

Then, the Schwinger integral over the particle-like light states in the Type IIB frame reads
\begin{equation}
\label{aFIIB}
a_8
=
\frac{2\pi r_1}{r_2}
\hat{\sum_{m_1,m_2}}
\int_0^\infty \frac{dt}{t}\,
e^{-\pi t \left( \frac{m_1^2}{r_1^2} + \frac{m_2^2}{r_2^2} \right)} .
\end{equation}
This integral has already been computed in \eqref{all0case}. Following the same steps and using as a regulator $4 \pi e^{- \gamma_E}\epsilon/({r_1r_2})$, we obtain
\begin{equation}
\label{aF}
    a_8
    = - \frac{2 \pi r_1}{r_{2}}
    \log\!\left(u \, \big|\eta (iu)\big|^4
    \right) .
\end{equation}
Although only one part of the one-loop contribution appears, the result is modular invariant thanks to the freedom in choosing the regulator. Comparing with \eqref{8dFull} we see that we obtain the full contribution for the expected instantons. However, as previously discussed, there is an extra instanton correction whose action lies within the window \eqref{window}. Specifically, $E\!D(-1)$-instantons have action $S_{E\!D(-1)}\sim1/g_s\sim1$,  and thus should be captured by the Schwinger integral over light states.

In the next section, we tackle the analogous F-theory limit in seven
dimensions and find that, indeed, all the instantons in the bound
\eqref{window} are obtained from an integral over the light states. In
other words, the absence of the $E\!D(-1)$-instanton contribution from
integrating out the light states is an artifact of working in eight
dimensions\footnote{A similar situation arose for the ten- and nine-dimensional $R^4$-term in the M-theory limit in \cite{Blumenhagen:2024ydy}, where also some contributions were missing. There, the discrepancy was attributed to a lack of particle states from wrapped $D2$-branes to generate all contributions. An analogous argument applies here: in the seven-dimensional F-theory limit, the light strings can wrap an additional direction which is absent in the eight-dimensional case, giving rise to light winding modes which generate the missing contributions.}. A heuristic reason for this behavior might be
that upon decompactification to 10D the elliptic fibration
of F-theory cannot have any degeneration loci and is
therefore also missing the $D7$-branes which are the magnetic
duals of $D(-1)$-branes.

There are two more limits of this type, denoted by F-th 2 and F-th 3 in Table~\ref{table_8d}. They are identical to the limit we have just discussed in the Type IIB frame after a series of dualities, namely after a T-duality along the first direction of the torus in the case of the limit F-th 2, or a T-duality along the second direction for the limit F-th 3. As a consequence, these other two limits exhibit the same content of light particle states and the corresponding Schwinger integral therefore coincides with that of the F-th~1 limit, giving again the coefficient \eqref{aF}.

\subsubsection{Decompactification/Decompactification* limit in 11D}

 Next, we consider the limit denoted by 11D in Table \ref{table_8d}
 which corresponds to a decompactification of three dimensions to
 M-theory. The light states correspond to KK modes along both
 directions of the torus together with the $D0$-branes. No light
 string appears in this limit. The induced instanton corrections are
 given by $E\!D0_{(i)}$-instantons and complex structure dependent terms with actions scaling as
\begin{equation}\label{instantons11D}
    S_{E\!D0_{(i)}} \sim \frac{m_{D0}}{m_{\mathrm{KK}_{(i)}}} \sim 1\,, \qquad
    S_{U_{(12)}} \sim \frac{m_{\mathrm{KK}_{(1)}}}{m_{\mathrm{KK}_{(2)}}} \sim 1\,.
\end{equation}
The Schwinger integral for these light states reads
\begin{equation}
\label{a11}
    a_8 = \frac{2 \pi}{t_{12}} 
    \hat {\sum_{m,m_1,m_2}}
    \int_0^\infty \frac{dt}{t} \,
    e^{-\pi t \left( \frac{m^2}{g_s^2} + \frac{m_1^2}{r_1^2} + \frac{m_2^2}{r_2^2} \right)} .
\end{equation}
Again, this integral has been dealt with in \eqref{a0all}. Following the same steps, we get 
\begin{equation}
\label{a11D}
    a_8
    =
    \frac{2\zeta(3)}{g_s^2}- \frac{2 \pi}{t_{12}} \log \!\left( u \, |\eta(iu)|^4 \right)
    + \frac{8 \pi}{g_s^2}
    \sum_{m>0}
    \hat{\sum_{m_1,m_2}} 
    \frac{m}{l}\,
    K_1(2 \pi m l)\, ,
\end{equation}
where now the regulator was chosen to be $4\pi e^{-\gamma_{\rm E}}t_{12}\epsilon$, so that a modular invariant combination appears. Comparing with \eqref{8dFull} we confirm that the full contributions to the instantons \eqref{instantons11D} appear.

This result could be interpreted as a ``1-loop'' M-theory contribution,
where the missing $2\pi^2/3$ term is the tree-level contribution and
Euclidean $E\!M2$-instantons provide the non-perturbative
corrections. The latter, i.e. worldsheet instantons in the Type IIA
frame, have action scaling as $S_{E\!F1_{(12)}}\sim
\lambda^{2}$, which is not within the proposed bound \eqref{window}. As we will see, the same conclusion applies to an analogous limit in the seven-dimensional setup, where the $E\!D2_{(123)}$ instantons are further suppressed and therefore do not contribute.

We now consider the limit denoted by 11D*, which is dual to the one
just discussed after performing a T-duality along both directions of
the torus. Thus, the light states are the winding modes along both
directions and wrapped $D2$-branes. 
These light states induce the same instanton corrections as the 11D limit, namely \eqref{instantons11D}. In this case, the Schwinger integral becomes
\begin{equation}
    a_8
    = \frac{2 \pi}{t_{12}}
    \hat{\sum_{n_{12},n_1,n_2}}
    \int_0^\infty \frac{dt}{t} \,
    e^{-\pi t \left(
        n_1^2 r_1^2
        + n_2^2 r_2^2
        + n_{12}^2 \frac{t_{12}^2}{g_s^2}
    \right)} .
\end{equation}
We have already dealt with this integral in the M-theory* limit in \eqref{aMcase1}. Therefore, by following exactly the same steps as in that case, one finds that the instanton expansion of $a_8$ is given by \eqref{a11D}, again confirming our proposal.

\subsubsection{Instanton expansion in the limits 10D/10D*}

Finally, we consider the decompactification limits to ten dimensions
labeled by 10D and 10D* in Table \ref{table_8d}. We start with the
former, in which the particle-like light states are KK modes along the
both directions of the torus together with $D0$-branes. There is also
a light fundamental string accompanying them. The particle-like light
content is the same as in the decompactification limit to eleven
dimensions, so that the induced instantons are given by
\eqref{instantons11D}. Moreover, the Schwinger integral takes the same
form as in ~\eqref{a11}. Therefore, following exactly the same steps
as in the eleven-dimensional decompactification limit, we confirm our
claim that the full instanton corrections with actions within the bound
\eqref{window} are obtained after integrating out just the light states in this limit.

We can now consider the dual limit 10D*, in which winding modes
along the 1 and 2 directions together with $D2$-branes wrapping both
directions are light. This limit is obtained from the 10D one by
applying T-duality along both directions of the torus. Again, the
particle-like content and their induced instanton corrections,
\eqref{instantons11D}, coincide with the dual eleven-dimensional
decompactification limit. Applying the same steps as in the last section we
obtain \eqref{a11D}, confirming our claim.

\subsection{Instanton expansion in generic limits}
\label{genericLimits}

So far we have only focused on special limits in moduli space, but of course it is fair to ask whether the bound \eqref{window} still captures the instanton corrections that are emergent in a generic infinite distance limit. To answer this question, we now reformulate it in terms of the taxonomy language of \cite{Etheredge:2024tok,Etheredge:2025ahf}. In this context, it is natural to extend the notion of $\alpha$-vector from \cite{Calderon-Infante:2020dhm} to the instanton $S$-vectors
\begin{equation}
\label{instantonvector}
    \vec S_i\coloneqq-\vec\nabla \log S_{i}\,,
\end{equation}
where $S_i$ is the action of the corresponding instanton in the canonically normalized moduli. In the conventions of \cite{Etheredge:2024tok}, in Appendix \ref{AppendixA} we have derived the eight-dimensional vectors to be given by
\eq{
\label{InstantonVectors}
    \vec{S}_{U_{(12)}}&=\left(\frac{1}{\sqrt{2}},0,-\sqrt{\frac{3}{2}}\right)\,,\quad\vec{S}_{E\!F1_{(12)}}=\left(0,{\sqrt{2}},0\right)\,,\\
    \vec{S}_{E\!D0_{(1)}}&=\left(-{\sqrt{2}},0,0\right)\,,\quad\quad\phantom{aa}\vec{S}_{E\!D0_{(2)}}=\left(-\frac{1}{\sqrt{2}},0,-\sqrt{\frac{3}{2}}\right)\,.
}
Although we have derived them from physical arguments, these vectors are in agreement with the taxonomy rules presented in \cite{Etheredge:2024amg, Etheredge:2025ahf}, which were argued to also include $\alpha$-vectors for instantons. Analogously to the tower vectors, given a direction in moduli space parametrized by some unit vector $\hat{u}$, the rate at which instanton actions become subleading is obtained by the projection $\vec S_i\cdot \hat u$. This allows us, due to the exponential nature of canonically normalized moduli, to reformulate the window \eqref{window} as
\begin{equation}\label{taxonomywindow}
    |\vec S_i\cdot \hat u|\leq (\vec\alpha_{\rm lightest}-\vec\Lambda_{\rm sp})\cdot \hat u\,,
\end{equation}
with $\vec\alpha_{\rm lightest}$ denoting the $\alpha$-vector of the lightest state\footnote{Note that the $\alpha$-vector for the lightest tower does not generically coincide with the $\alpha$-vector belonging to the tower polytope along the same direction. The $\alpha$-vector for the lightest tower lies in what is referred to in \cite{Etheredge:2023odp} as the Max-$\alpha$ hull. For a direction $\hat u$ it is given by
\begin{equation*}
    \vec\alpha_{\rm lightest}=(\max_{i\in I} \{\vec\alpha_i\cdot\hat u\})\,\hat u\,,\quad I=\{\text{particle states}\}\,.
\end{equation*} } along the direction $\hat u$ and $\vec\Lambda_{\rm sp}$ the associated species vector\footnote{Again one needs to use the maximum of the projections of the species vectors associated to each tower, similar to the case of $\vec\alpha_{\rm lightest}$.} along the same direction. 

Thus, we can plot the upper bound \eqref{taxonomywindow} in every direction of the moduli space, obtaining a closed surface. Furthermore, by also plotting the projections of the instanton actions in every direction, namely $|\vec S_i\cdot \hat u|\,\hat u$, we derive a graphical criterion of which instanton corrections get generated by the Schwinger integral over the light states. 
As in the special cases studied above, by generating the instanton
corrections we really mean producing the \textit{total} contribution
to such an instanton that appears in the full amplitude 
\eqref{8dFull}.

\begin{figure}[ht]
\centering
\includegraphics[width=\linewidth]{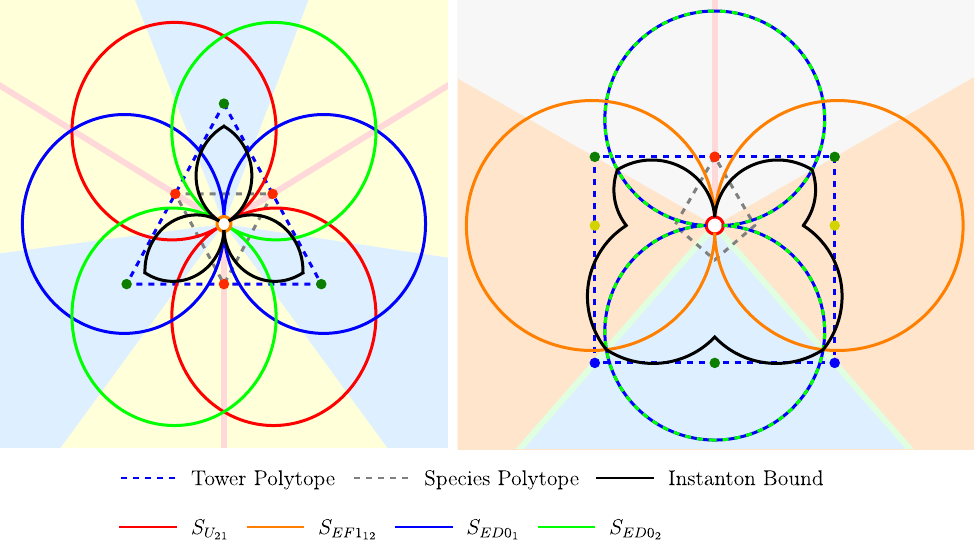}
\caption{This figure shows the two slices of the tower and species polytopes in Figures \ref{taxonomyTower8d} and \ref{taxonomySpecies8d} displayed on top of another. We also plot each of the instanton actions projected along both slices and the bound \eqref{taxonomywindow}. The shaded areas correspond to regions of moduli space with the same instanton corrections.}
\label{instantonSlices8d}
\end{figure}

In Figure \ref{instantonSlices8d} we have plotted the bound and the projections of the instanton actions for the two slices of Figures \ref{taxonomyTower8d} and \ref{taxonomySpecies8d}. First, we note that in emergent string limits the bound \eqref{taxonomywindow} degenerates to zero, with the graphical depiction offering little insight. Moreover, along the first (second) slice the worldsheet instantons (complex structure) have null projection, again spoiling the graphical proposal. To remedy this, we have added a small radius $\epsilon$ to every projection. 

Since the figures are quite convoluted, let us provide an example of how they can be utilized to infer the emergent instanton content along a general direction. Consider the decompactification limit to eleven dimensions, given by the point labeled with a $3$ in the rightmost edge in the second figure of Figure \ref{taxonomyTower8d}, which we have studied already in the previous section. Looking now at the analogue in Figure \ref{instantonSlices8d}, along that direction, the instanton bound encloses the complex structure, $E\!D0_{(1)}$-instantons and $E\!D0_{(2)}$-instantons, while the worldsheet instantons are outside the bound and thus should not be obtained from an emergence computation. Indeed, as shown in \eqref{a11D} this is precisely the instanton content that gets generated in this limit\footnote{The polynomial tree-level term $2\zeta(3)/g_s^2$ also gets generated. Our proposal does not predict which of such polynomial terms are generated, we leave this question open for the moment.}. 

Regarding the rest of the infinite distance limits visualized in these polytopes, we have divided the slices into different colored regions according to the instanton content. The predicted instanton content for each region reads
\setlength{\fboxsep}{2pt}
\eq{
    a_{8\,,\,\fcolorbox{black}{lightyellow}{$\scriptscriptstyle \phantom{I}$}}&\sim -\frac{2\pi}{t_{12}}\log\left(t_{12}|\eta(it_{12})|^4\right)\,,
    \\ a_{8\,,\,\fcolorbox{black}{lightblue}{$\scriptscriptstyle \phantom{I}$}}&\sim -\frac{2\pi r_1'}{r_2'}\log\left(r_2'^2|\eta(it_{12}')\eta(iu')|^4\right)\,,
    \\ a_{8\,,\,\fcolorbox{black}{lightgray}{$\scriptscriptstyle \phantom{I}$}}&\sim -\frac{2\pi}{t_{12}}\log\left(u|\eta(iu)|^4\right)\,,
    \\ a_{8\,,\,\fcolorbox{black}{lightorange}{$\scriptscriptstyle \phantom{I}$}}&\sim - \frac{2 \pi}{t_{12}} \log \!\left( u \, |\eta(iu)|^4 \right)
    + \frac{8 \pi}{g_s^2}\sum_{m>0} \hat{\sum_{m_1,m_2}}\frac{m}{l}\,K_1(2 \pi m l)\,.
}
In the blue regions we have expressed the instanton contribution in
the more appropriate Type IIB frame. The remaining shaded regions, the light red $\fcolorbox{black}{lightred}{\phantom{a}}$ and light green $\fcolorbox{black}{lightgreen}{\phantom{a}}$ areas, correspond to the emergent string limits and the M-theory limits, respectively, which were already studied in the previous sections.

What remains to be confirmed is that, indeed, the light states are
enough to generate these contributions. Instead of looking at each
region and confirming this, let us note that the Schwinger integrals
performed until now are enough to show this after checking what towers
of states are below the species scale\footnote{The
  hierarchy of mass scales can be obtained in the taxonomy
  description by following the recipe presented in Section
  \ref{taxonomySection}.}.  Similarly to what happened in the F-theory
limits, in the light blue shaded regions the light particle content is
not enough to generate both of the expected instanton contributions (in the Type IIB frame).
Nonetheless, we believe that this is again an artifact
of the less generic setup in eight dimensions  and  that this issue will not
arise in the lower-dimensional moduli spaces.

Let us emphasize that our analysis has confirmed that the M-theory limit of
\cite{Blumenhagen:2024ydy} remains as the only limit in which the
entire $R^4$-coupling can arise from a single Schwinger integral summing
over the light towers of states. This clearly singles out this special
limit and is consistent with  the M-theoretic Emergence Proposal.

\section{Instanton contributions to the \texorpdfstring{$R^4$}{R⁴}-term in 7D}
\label{section4}
Having confirmed our claim in eight dimensions, one might wonder
whether the light particle content in these infinite-distance limits
was overly restrictive. For that reason, we study several more
involved limits in seven dimensions and confirm that integrating out
the light states reproduces the full instanton contribution in the
window \eqref{window},
while the heavy state contributions are redundant.

Let us remark that although the contributions from integrating
out states in seven dimensions are relatively straightforward, the
fact that in lower dimensions the light states provide the
\textit{complete} contribution associated with the instantons in the
bound \eqref{window} is highly non-trivial. Assuming the validity of
our proposal, the graphical approach we have just developed would, in
principle, allow us to explore the full moduli space of the
lower-dimensional theories to determine the generic patterns that emerge.

\subsection{The M-theory limit}
In seven dimensions the M-theory limit is defined via the scaling of the Type IIA parameters
\eq{
       g_s\to \lambda^{3} g_s
       \,,\qquad r_i\to \lambda r_i\,,\qquad M_s\to
       \lambda^\frac{3}{5}  M_s
     }
with  $\lambda\to\infty$, such that the seven-dimensional Planck mass remains constant. Changing to M-theory parameters \eqref{mparameters}, the radii of the three directions of $T^3$ are kept fixed in M-theory units, i.e. $M_* R_i$ does not scale. In this limit the lightest tower of states, as in the case of eight dimensions, are the $D0$-branes of mass scale $M_{D0}\sim  {M_s/g_s}\sim\lambda^{-\frac{12}{5}}$, which induce a species scale
\eq{
         \Lambda_{\rm sp}\sim M_{\rm Pl}^\frac{5}{6} M_{D0}^\frac{1}{6}\sim  \lambda^{-\frac{2}{5}}\,,
       }
as follows from the general formula given in \eqref{speciesscale}.

Scanning over all possible towers of particle-like states in 7D, one
finds the list of light towers shown in Table \ref{table_Mtheory}.
%%%%%%%%%%%%%%%%%%%%%%%%%
\begin{table}[ht] 
\renewcommand{\arraystretch}{1.2} 
\begin{center} 
\begin{tabular}{|l|l||l|l|} 
\hline
light towers    & mass scale & instantons & action   \\
\hline \hline
$D0$  &   $M\sim \frac{M_s}{g_s}\sim\lambda ^{-\frac{12}{5}}$ &  $U_{(ij)}$ &
 $S\sim \frac{m_{{\rm KK}_{(j)}}}{m_{{\rm KK}_{(i)}}}\sim \frac{m_{D2_{(ik)}}}{m_{D2_{(jk)}}}\sim 1$ \\
$D2_{(ij)}$  &   $M\sim \frac{M_s}{g_s} r_i r_j\sim \lambda^{-\frac{2}{5}}$ & $E\!D0_{(i)}$
                          & $S\sim \frac{m_{D0}}{m_{{\rm KK}_{(i)}}}\sim \lambda^{-2}$ \\
${\rm KK}_{(i)}$  &   $M\sim \frac{M_s}{r_i}\sim \lambda ^{-\frac{2}{5}}$ &    $E\!F1_{(ij)}$ &  $S\sim \frac{m_{D2_{(ij)}}}{m_{D0}}\sim \lambda^{2}$ \\
 &    &   $E\!D2_{(ijk)}$  &   $S\sim \frac{m_{D2_{(ij)}}}{m_{{\rm KK}_{(k)}}}\sim 1$ \\    
  \hline\hline
$D2_{(i)}$  &   $T\sim \frac{M_s^2 r_i}{g_s}\sim\lambda
             ^{-\frac{4}{5}}$ &   &  \\
  \hline
\end{tabular}
\caption{Mass scales of light towers of particles/strings and their
  induced instanton corrections in the M-theory decompactification limit.}
 \label{table_Mtheory}
\end{center} 
\end{table}
Note that the six combinations $D2_{(ij)}-{\rm KK}_{i}$ with KK
momentum along a D2-brane are not BPS and hence their mass ratios do
not appear in  the right column of Table \ref{table_Mtheory}. Moreover, the three particle-like states from wrapped $F1$-strings are heavy in this limit and are thus treated as non-perturbative. Once again, although these contributions are not directly visible in the Schwinger integral, the seven light towers of states in Table~\ref{table_Mtheory} capture {\it all} potential instantons with actions in the window~\eqref{window}, just as in the eight-dimensional case.

By successively splitting the sums and performing Poisson resummations, the Schwinger integral for the light towers was evaluated explicitly in \cite{Blumenhagen:2024ydy}. For later purpose let us recall the result. 
To present the results in a compact form, we introduce a vector notation to distinguish the different classes of solutions to the BPS condition. In seven dimensions, given
\eq{
    S = \Big\{ 
    (1,0,0), (0,1,0), (0,0,1), 
    (1,1,0), (1,0,1), (0,1,1), 
    (1,1,1)
    \Big\} \,,
}
then the Schwinger integral over the light states is found to be
\begin{empheq}[box=\fbox]{align}
  \label{mtheoryschwingerresults}
  \  a_7&= \frac{2\pi^2}{3} + \sum_{\alpha\in S}  {\cal J}^{\alpha}\Big(\frac{t_{123}}{g_s};t_{12} , t_{13}, t_{23}\Big) \nonumber\\
  \  &+\frac{2\zeta(3)}{g_s^2}+
  \frac{8\pi}{g_s^2} \sum_{m>0} \hat{\sum_{m_1,m_2,m_3}} \frac{m}{l}
    K_1\big(2\pi  m l\big)
  + \frac{2\pi}{t_{123}}{\cal E}^{\rm SL(3)}_{\mathbf{3};s=1/2}\Big(\frac{1}{r_1},\frac{1}{r_2},\frac{1}{r_3}\Big)\,.
  \end{empheq}
Here, we have defined $t_{123}=r_1r_2r_3$, $t_{ij}=r_i r_j$ and
\eq{
       {\cal J}^{\alpha}(M_0;M_1,M_2,M_3)  =2\pi \,2^{|\alpha|} \sum_{n>0}
       \sum_{\mathbf{\tilde n}_\alpha>0} \sum_{(m,p)\ne (0,0)}
       \frac{e^{-2\pi n L_\alpha }}{L_\alpha }\,,
}
with
\eq{
       L_\alpha=
       \sqrt{p^2 M_0^2+   {\textstyle \sum_{i=1}^{3} } \alpha_i m^2
         \tilde n^2_i M^2_i}\,.
}
The first line of \eqref{mtheoryschwingerresults} is interpreted as the contribution from bound states of $E\!D2$ and up to three worldsheet instantons $E\!F1_{ij}$, while
the second line contains the tree level term and introducing
\eq{
           l= \frac{1}{g_s} \sqrt{m_1^2 r_1^2+m_2^2 r_2^2 + m_3^2 r_3^2}\,,
}
one recognizes the action of the $E\!D0$-instanton contributions. The last term in the second line is the Eisenstein series
\eq{ 
  {\cal E}^{\rm SL(3)}_{\mathbf{3};s=1/2}(M_1,M_2,M_3)=\hat{\sum_{n_1,n_2,n_3}}\int_0^\infty
  \frac{dt}{t^\frac{1}{2}} \, e^{-\pi t \big(\sum_i M_i^2 n_i^2\big)}
}
which does not involve any further BPS constraint and can be evaluated to give a complex structure dependent term \cite{Blumenhagen:2024ydy}. 

As in 8D,  we have found that  integrating out the
light towers of states yields all potential instanton corrections.
Even though the full computation is cumbersome, we believe that  integrating out in the total $1/2$-BPS Schwinger integral also
the heavy towers of wrapped fundamental strings will  not change the
result \eqref{mtheoryschwingerresults} and hence be redundant.
Having determined
the complete seven-dimensional coupling, we now consider several
notable infinite-distance limits.

\subsection{The  emergent perturbative Type IIA string}
First, we consider the straightforward weakly coupled Type IIA string limit. This  is defined via the infinite distance limit
\eq{
  \label{typeIIAstringiialimit}
       g_s\to \lambda g_s \,,\qquad r_i\to r_i\,,\qquad M_s\to
       \lambda^\frac{2}{5}  M_s
     }
with $i=1,2,3$ and  $\lambda\to 0$.

Scanning through the list towers of 1/2-BPS  of states \eqref{massBPSstates}, the ones shown in Table \ref{table_emergeIIAweak} are light, i.e. have mass scales parametrically not exceeding the species scale, which is $\Lambda_{\rm sp}\sim \lambda ^\frac{2}{5}$.
These correspond to the light towers of weakly coupled Type IIA string theory, all having parametrically the same mass equal to the species scale, $\Lambda_{\rm sp}=M_s$.

%%%%%%%%%%%%%%%%%%%%%%%%%
\begin{table}[ht] 
\renewcommand{\arraystretch}{1.2} 
\begin{center} 
\begin{tabular}{|l|l||l|l|} 
\hline
light towers    & mass scale & instantons & action   \\
\hline \hline
$F1_{(i)}$  &   $M\sim M_sr_i\sim \lambda^\frac{2}{5}$ & $E\!F1_{(ij)}$
                          & $S\sim \frac{m_{F1_{(i)}}}{m_{{\rm
                            KK}_{(j)}}}\sim \frac{m_{F1_{(j)}}}{m_{{\rm KK}_{(i)}}}\sim 1$ \\
${\rm KK}_{(i)}$  &   $M\sim \frac{M_s}{r_i}\sim \lambda ^{\frac{2}{5}}$ &    $U_{(ij)}$ &  $S\sim  \frac{m_{{\rm KK}_{(j)}}}{m_{{\rm
                                       KK}_{(i)}}} \sim \frac{m_{F1_{(i)}}}{m_{F1_{(j)}}}\sim 1$ \\
  \hline\hline
$F1$  &   $T\sim {M_s^2 }\sim\lambda
              ^{\frac{4}{5}}$ &   &  \\
  \hline
\end{tabular}
\caption{Mass scales of light towers of particles/strings and their
  induced instanton corrections for the weakly coupled Type IIA string.}
 \label{table_emergeIIAweak}
\end{center} 
\end{table}

Let us now show explicitly that integrating out the light towers of states already reproduces the full contribution of the worldsheet instantons and the $U$-moduli, as in the eight-dimensional case. Since the complete $R^4$-term in seven dimensions is known from the M-theory limit, we can directly check whether including the heavy towers in the full $1/2$-BPS Schwinger integral \eqref{r47dschwinger} yields only a trivial contribution to the Type IIA one-loop term.

To proceed, let us first evaluate the Schwinger integral for the light towers of states in the Type IIA frame, which takes the form
\eq{
\label{stringschwinger}
a_{7}\simeq \frac{2\pi}{r_1 r_2 r_3} \hat{\sum_{m_i,n_i}} \int_0^\infty  \frac{d t}{t^{\frac12}}\;
\delta({\rm BPS})\, e^{-{\pi t}\sum_i\big(
     \frac{m_i^2}{r_i^2} +  n_i^2 r_i^2\big)}\,,
}
with the BPS constraint
\eq{
  n_1 \,m_1 + n_2\, m_2+ n_3\, m_3=0\,.
}  
Now the Schwinger integral  \eqref{stringschwinger} can be evaluated completely analogous to the computation performed in Appendix A of \cite{Blumenhagen:2024ydy}, leading to the result
\eq{
  \label{IIAstringschwingerresults}
  a_7=\frac{2\pi^2}{3} + \sum_{\alpha\in S}  {\cal I}^{\alpha}\Big(t_{12},t_{13},t_{23}\Big)
  + \frac{2\pi }{t_{123}}\, {\cal E}^{\rm
    SL(3)}_{\mathbf{3};s=1/2}\Big(\frac{1}{r_1},\frac{1}{r_2},\frac{1}{r_3}\Big)
}
with
\eq{
\label{calIalpha}
       {\cal I}^{\alpha}(M_1,M_2,M_3)  =4\pi \,2^{|\alpha|} \sum_{n>0} \sum_{m_\alpha>0}  \frac{e^{-2\pi n
           L_\alpha }}{L_\alpha }\,
 }
and     
\eq{
\label{Lalpha}
          L_\alpha=
          \sqrt{{\textstyle \sum_{i=1}^{3} } \alpha_i m^2_i M^2_i}\,.
}          
This is to be compared with \eqref{mtheoryschwingerresults}. The constant $2\pi^2/3$ and the Eisenstein series terms clearly agree, while the remaining two terms in the second line of \eqref{mtheoryschwingerresults} correspond to the Type IIA tree-level and $E\!D0$-instanton contributions.
Hence, it remains to 
evaluate ${\cal J}^{\alpha}$ for vanishing coefficients
of the $E\!D2_{(123)}$-instanton.
Indeed, utilizing the relation \eqref{sumcoprimes} one can show   that
\eq{
  \label{hannover96}
            \sum_{\alpha\in S}  {\cal J}^{\alpha}\Big( \frac{t_{123}}{g_s}; t_{12}, t_{13}, t_{23}\Big)\Big\vert_{{\rm no}\;E\!D2_{(123)}}=
            \sum_{\alpha\in S}  {\cal I}^{\alpha}\Big( t_{12},t_{13}, t_{23}\Big)\,.
}
This shows that integrating out the light towers of states in the Type IIA emergent string limit already yields the full one-loop amplitude. When the heavy towers are included, their contributions cancel non-trivially among themselves and thus vanish.

\subsection{The emergent Type IIB string}
Next, we study  the seven-dimensional analogue of the ESB limit of
Table \ref{table_8d}.

\subsubsection{The Type IIA limit}

In this case, the limit in the Type IIA frame is defined as follows:
\eq{
  \label{emstringiialimit}
       g_s\to \lambda g_s\,,\qquad r_1\to \lambda^{-1}
       r_1
       \,,\qquad r_j\to \lambda  r_j\,,\qquad M_s\to
       \lambda^\frac{1}{5}  M_s
     }
with $j=2,3$ so that the seven dimensional Planck scale is indeed kept fixed for $\lambda\to\infty$.

Scanning through the list of towers of 1/2-BPS  states \eqref{massBPSstates}, the ones shown in Table \ref{table_emergeIIB} are light, i.e. have mass scales parametrically not exceeding the species scale, which is $\Lambda_{\rm sp}\sim \lambda ^{-\frac{4}{5}}$.
%%%%%%%%%%%%%%%%%%%%%%%%%
\begin{table}[ht] 
\renewcommand{\arraystretch}{1.2} 
\begin{center} 
\begin{tabular}{|l |l||l|l|} 
\hline
light towers    & mass scale & instantons & action   \\
\hline \hline
$D0$  &   $M\sim \frac{M_s}{g_s}\sim\lambda ^{-\frac{4}{5}}$ &  $U_{(23)}$ &
 $S\sim  \frac{m_{{\rm KK}_{(3)}}}{m_{{\rm KK}_{(2)}}}\sim  \frac{m_{D2_{(12)}}}{m_{D2_{(13)}}}\sim 1$ \\
$D2_{(1j)}$  &   $M\sim \frac{M_s}{g_s} r_1 r_j\sim \lambda^{-\frac{4}{5}}$ & $E\!D0_{(j)}$
                          & $S\sim \frac{m_{D0}}{m_{{\rm
                            KK}_{(j)}}}\sim \frac{m_{D2_{(1j)}}}{m_{F1_{(1)}}}\sim 1$ \\
${\rm KK}_{(j)}$  &   $M\sim \frac{M_s}{r_j}\sim \lambda ^{-\frac{4}{5}}$ &    $E\!F1_{(1j)}$ &  $S\sim  \frac{m_{F1_{(1)}}}{m_{{\rm
                                       KK}_{(j)}}} \sim \frac{m_{D2_{(1j)}}}{m_{D0}}\sim 1$ \\
$F1_{(1)}$  &   $M\sim {M_s r_1}\sim \lambda ^{-\frac{4}{5}}$ &   $E\!D2_{(123)}$  &   $S\sim \frac{m_{D2_{(12)}}}{m_{{\rm KK}_{(3)}}}\sim  \frac{m_{D2_{(13)}}}{m_{{\rm KK}_{(2)}}}\sim 1$ \\    
  \hline\hline
$D2_{(1)}$  &   $T\sim \frac{M_s^2 r_1}{g_s}\sim\lambda
              ^{-\frac{8}{5}}$ &   &  \\
  \hline
\end{tabular}
\caption{Mass scales of light towers of particles/strings and their induced instanton corrections for an emergent string limit in Type IIA frame.}
\label{table_emergeIIB}
\end{center} 
\end{table}

\noindent

Apparently, Table $\ref{table_emergeIIB}$ shows all the features
expected for an emergent string limit, with  the single light string
being the $D2$-brane wrapped on the seventh direction. The two right columns list the induced instantons and their actions for all $1/2$-BPS combinations of two light towers of states. The combinations $D2_{(1j)}-{\rm KK}_{(j)}$ and $D0-F1_{(1)}$ are not BPS and therefore do not appear in Table \ref{table_emergeIIB}. Integrating out the light towers misses the four instantons
\eq{
S_{U_{(1j)}}\sim \frac{r_1}{r_j}\sim \lambda^{2}\,,\qquad
S_{E\!D0_{(1)}}\sim \frac{r_1}{g_s}\sim \lambda^{-2}\,,\qquad
S_{E\!F1_{(23)}}\sim r_2 r_3\sim \lambda^{2}\,,
}
whose actions lie outside the window \eqref{window}.

\subsubsection{The dual Type IIB limit}
This emergent string limit is better understood after applying a T-duality in the first direction followed by an S-duality. This duality chain acts on the moduli parameters as in \eqref{typeIIBmoduli}, with $r_3$ transforming in the same way as $r_2$. Consequently, the limit~\eqref{emstringiialimit} becomes
\eq{
       g_s'\to \lambda^{-2} g_s'\,,\qquad M_s'\to
       \lambda^{-\frac{4}{5}} M_s'\,,\qquad r_i'\to r_i'\,.
}            
Then the light towers of states and the corresponding instantons are one-to-one mapped as shown in Table \ref{table_emergeIIBb}.

%%%%%%%%%%%%%%%%%%%%%%%%%
\begin{table}[ht] 
\renewcommand{\arraystretch}{1.2} 
\begin{center} 
\begin{tabular}{|l|l||l|l|} 
\hline
light towers    & mass scale & instantons & action   \\
\hline \hline
$F1_{(1)}$  &   $M\sim {M_sr_1}\sim\lambda ^{-\frac{4}{5}}$ &  $U_{(23)}$ &
 $S\sim  \frac{m_{{\rm KK}_{(3)}}}{m_{{\rm KK}_{(2)}}}\sim  \frac{m_{F1_{(2)}}}{m_{F1_{(3)}}}\sim 1$ \\
$F1_{(j)}$  &   $M\sim M_sr_j\sim \lambda^{-\frac{4}{5}}$ & $E\!F1_{(1j)}$
                          & $S\sim \frac{m_{F1_{(1)}}}{m_{{\rm
                            KK}_{(j)}}}\sim \frac{m_{F1_{(j)}}}
                                       {m_{{\rm KK}_{(1)}}}\sim 1$ \\
${\rm KK}_{(j)}$  &   $M\sim \frac{M_s}{r_j}\sim \lambda ^{-\frac{4}{
              5}}$ &    $U_{(1j)}$ &  $S\sim  \frac{m_{{\rm KK}_{(j)}}}{m_{{\rm
                                       KK}_{(1)}}} \sim \frac{m_{F1_{(1)}}}
                                       {m_{F1_{(j)}}}\sim 1$ \\
${\rm KK}_{(1)}$  &   $M\sim \frac{M_s}{r_1}\sim \lambda ^{-\frac{4}
              {5}}$ &   $E\!F1_{(23)}$  &   $S\sim \frac{m_{F1_{(2)}}}
                                        {m_{{\rm KK}_{(3)}}}\sim  \frac{m_{F1_{(3)}}} {m_{{\rm KK}_{(2)}}}\sim 1$ \\    
  \hline\hline
$F1$  &   $T\sim {M_s^2 }\sim\lambda
              ^{-\frac{8}{5}}$ &   &  \\
  \hline
\end{tabular}
\caption{Mass scales of light towers of particles/strings and their
  induced instanton corrections for dual weakly coupled Type IIB string.}
 \label{table_emergeIIBb}
\end{center} 
\end{table}

This corresponds to the weakly coupled Type IIB string limit, where the light towers of states are the fundamental string excitations together with their KK and winding modes along the three toroidal directions. The species scale coincides with the string scale, $\Lambda_{\rm sp}=M_s$, while the $D$-brane towers are heavier. The actions of the corresponding instantons are parametrically of order one, $S_{\rm inst}\sim 1$, in agreement with \eqref{window}, since the species scale equals the mass scale of the light towers. In contrast, the actions of $D$-brane instantons, such as $E\!D(-1)$ or wrapped $E\!D1$-branes, scale as
\eq{
   S_{E\!D(-1)}\sim \frac{1}{g_s}\sim \lambda^{2}\,,\qquad
  S_{E\!D1_{(ij)}}\sim \frac{r_i r_j}{g_s}\sim \lambda^{2}
}
and are therefore parametrically larger than one.

\subsubsection{Instanton expansion}

We now show that integrating out the light towers of states already reproduces the full contribution of the worldsheet instantons and $U$-moduli. To this end, we first evaluate the Schwinger integral for the light towers in the Type IIB frame, which takes the form
\eq{
\label{stringschwingerIIB}
a_{7}\simeq \frac{2\pi r_1}{r_2 r_3} \hat{\sum_{m_i,n_i}} \int_0^\infty  \frac{d t}{t^{\frac12}}\;
\delta({\rm BPS})\, e^{-{\pi t}\sum_i\big(
     \frac{m_i^2}{r_i^2} +  n_i^2 r_i^2\big)}\,,
}
with the BPS constraint
\eq{
  m_1\,n_1 + n_2\,m_2+ n_3\,m_3=0\,.
}  

Now the Schwinger integral \eqref{stringschwingerIIB} can be evaluated identically to the previous weakly coupled Type IIA limit. The only difference is that we fix the integers $\{m_1, n_2, n_3\}$ and solve the Diophantine equation for $\{n_1, m_2, m_3\}$, similar to the procedure presented in eight dimensions. Then the result can be expressed as
\eq{
  \label{stringschwingerresults}
  a_7=\frac{2\pi^2}{3} + \sum_{\alpha\in S}  {\cal I}^{\alpha}\Big(t_{23},\frac{r_2}{r_1},\frac{r_3}{r_1}\Big)
  + \frac{2\pi r_1}{r_2 r_3}\, {\cal E}^{\rm SL(3)}_{\mathbf{3};s=1/2}\Big(r_1,\frac{1}{r_2},\frac{1}{r_3}\Big)\,,
}
where $\mathcal{I}^\alpha$ and $L_\alpha$ are given by \eqref{calIalpha} and \eqref{Lalpha}, respectively.

This is to be compared with the total coefficient
\eqref{mtheoryschwingerresults} after transforming to the Type IIB frame via the chain of dualities presented above. We
obtain
\eq{
  \label{mtheorychwingerresultstypeIIB}
  a_7=&\frac{2\pi^2}{3} + \sum_{\alpha\in S}  {\cal J}^{\alpha}\Big( t_{23};\frac{r_2}{r_1}, \frac{r_3}{r_1}, \frac{t_{23}}{g_s}\Big)\\
  &+\frac{2 r_1^2\zeta(3)}{g_s^2}+
  \frac{8\pi r_1^2}{g_s^2} \sum_{m>0} \hat{\sum_{m_1,m_2,m_3}} \frac{m}{l}
    K_1\big(2\pi  m l\big)
  + \frac{2\pi r_1}{r_2 r_3}\,{\cal E}^{\rm SL(3)}_{\mathbf{3};s=1/2}\Big(r_1,\frac{1}{r_2},\frac{1}{r_3}\Big)
}
with
\eq{\label{linIIB}
          l= \sqrt{ \frac{m_1^2}{g_s^2} +m_2^2\Big( \frac{t_{12}}{g_s}\Big)^2 + +m_3^2\Big( \frac{t_{13}}{g_s}\Big)^2}\,.
}
This contains both perturbative contributions in the Type IIB string coupling $g_s$ and non-perturbative terms in an intertwined manner. Again, the terms involving the Eisenstein series in \eqref{stringschwingerresults} are in clear agreement with those in the full expression \eqref{mtheorychwingerresultstypeIIB}. The remaining two terms in the second line of \eqref{mtheorychwingerresultstypeIIB} can be identified as the Type IIB tree-level contribution and the $E\!D(-1)$ and $E\!D1_{(1j)}$-instanton contributions, respectively.

Furthermore, by employing the natural generalization of \eqref{hannover96}, one finds that integrating out only the light towers of states in this Type IIB emergent string limit is sufficient to reproduce the full one-loop amplitude.

\subsection{An F-theory limit}

Let us next investigate the seven-dimensional version of the F-theory
limit, which is a decompactification limit to nine-dimensions.
We now verify our earlier claim that, in this case, the full content
of instanton corrections in the window~\eqref{window} is captured, in
contrast to the eight-dimensional case, where certain instanton
contributions were absent.

\subsubsection{The Type IIA limit}

The scaling of the Type IIA parameters reads
\eq{
  \label{ftheorylimit}
       g_s\to \lambda g_s\,,\qquad r_1\to \lambda^{-1}
       r_1
       \,,\qquad r_j\to r_j\,,\qquad M_s\to
       \lambda^\frac{3}{5}  M_s
     }
 with $j=2,3$  and $\lambda\to\infty$.
 It turns out that there are two towers of light particle modes of  degenerate mass scale $M_{\rm light}\sim \lambda^{-\frac{7}{5}}$,  arising from  $D2$-branes wrapped around the planes $(12)$ and $(13)$.  The induced species scale  is computed via \eqref{speciesscale}
 \eq{
         \Lambda_{\rm sp}\sim M_{\rm Pl}^\frac{5}{7} \, M_{\rm light}^\frac{2}{7}\sim  \lambda^{-\frac{2}{5}}\,.
       }
Scanning through all towers of BPS states, the present light towers and their induced instanton corrections are shown in Table \ref{table_emergeF}.

%%%%%%%%%%%%%%%%%%%%%%%%%
\begin{table}[ht] 
\renewcommand{\arraystretch}{1.2} 
\begin{center} 
\begin{tabular}{|l|l||l|l|} 
\hline
light towers    & mass scale & instantons & action   \\
  \hline \hline
  $D2_{(1j)}$  &   $M\sim \frac{M_s}{g_s} r_1 r_j\sim \lambda^{-\frac{7}{5}}$ & $U_{(1j)}$
                          & $S\sim \frac{m_{D2_{(1k)}}}{
                            m_{D2_{(kj)}}}\sim  \lambda^{-1}$ \\
$D0$  &   $M\sim \frac{M_s}{g_s}\sim\lambda ^{-\frac{2}{5}}$ &  $U_{(23)}$ &
 $S\sim   \frac{m_{D2_{(12)}}}{m_{D2_{(13)}}}\sim 1$ \\
$D2_{(23)}$  &   $M\sim \frac{M_s}{g_s} r_2 r_3\sim \lambda^{-\frac{2}{5}}$ & $E\!D0_{(j)}$
                          & $S\sim \frac{m_{D2_{(1j)}}}{
                                       m_{F1_{(1)}}}\sim  \lambda^{-1}$ \\
$F1_{(1)}$  &   $M\sim {M_s r_1}\sim \lambda ^{-\frac{2}{
              5}}$ & $E\!F1_{(1j)}$ &  $S \sim \frac{m_{D2_{(1j)}}}{
                                    m_{D0}}\sim  \lambda^{-1}$ \\
    &   & $E\!F1_{(23)}$ &  $S \sim \frac{m_{D2_{(23)}}}{
                                       m_{D0}}\sim 1$ \\   
  \hline\hline
$D2_{(1)}$  &   $T\sim \frac{M_s^2 r_1}{g_s}\sim\lambda
              ^{-\frac{4}{5}}$ &   &  \\
  $D4_{(123)}$  &   $T\sim \frac{M_s^2 r_1r_2 r_3}{g_s}\sim\lambda
              ^{-\frac{4}{5}}$ &   &  \\
  \hline
\end{tabular}
\caption{Mass scales of light towers of particles/strings and their
  induced instanton corrections for an F-theory like  decompactification limit.}
 \label{table_emergeF}
\end{center} 
\end{table}

One notices that there are two types of strings whose mass scales are parametrically of the same order as the species scale, a characteristic feature of the F-theory limit. Moreover, the
combinations $D0-F1_{(1)}$ and $D2_{(23)}-F1_{(1)}$ are not BPS and
hence their mass ratios do not appear in the right columns of Table \ref{table_emergeF}. Since all instanton actions lie in the window \eqref{window}
one is missing the two instantons
\eq{
  S_{E\!D0_{(1)}}\sim \frac{r_1}{g_s}\sim \lambda^{2}\,,\qquad
  S_{E\!D2_{(123)}}\sim \frac{t_{123}}{g_s}\sim \lambda^{2}\ ,
} 
which apparently have actions outside the window \eqref{window}. According to our philosophy these should be considered as extra genuinely non-perturbative contributions in this decompactification limit.

\subsubsection{The dual Type IIB limit}

By applying a chain of T- and S-dualities, the latter limit can be mapped to a Type IIB limit, where the decompactification to 9D is manifest. For that purpose one first performs a T-duality along the first direction, followed by an S-duality and then two more T-dualities along
the second and  third directions. In total this acts as
\eq{
        g_s'=\frac{1}{r_2 r_3}\,,\qquad M_s'=M_s \sqrt{\frac{r_1}{
          g_s}}\,,\qquad
         r_1'=\frac{1}{\sqrt{g_s r_1}}\,,\qquad r_j'=\frac{\sqrt{g_s}}{\sqrt{r_1} r_j }
      }
so that the limit \eqref{ftheorylimit} becomes
\eq{
       g_s'\to g_s'\,,\qquad M_s'\to
       \lambda^{-{\frac{2}{5}}} M_s'\,,\qquad r_1'\to r_1'\,,\qquad
       r_j'\to \lambda\, r_j'\,.
    }            
Hence, the string coupling is not scaling at all and is thought to be of order one. In the Type IIB frame the light towers of states and corresponding instanton contributions are shown in Table \ref{table_emergeFb}.

%%%%%%%%%%%%%%%%%%%%%%%%%
\begin{table}[ht] 
\renewcommand{\arraystretch}{1.2} 
\begin{center} 
\begin{tabular}{|l|l||l|l|} 
\hline
light towers    & mass scale & instantons & action   \\
  \hline \hline
  ${\rm KK}_{(j)}$  &   $M\sim \frac{M_s}{r_j} \sim \lambda^{-\frac{7}{5}}$ & $E\!D1_{(1j)}$
                          & $S\sim \frac{m_{D1_{(1)}}}{m_{{\rm KK}_{(j)}}}\sim  \lambda$ \\
$F1_{(1)}$  &   $M\sim M_s r_1\sim\lambda ^{-\frac{2}{5}}$ &  $U_{(23)}$ &
 $S\sim   \frac{m_{{\rm KK}_{(3)}}}{m_{{\rm KK}_{(2)}}}\sim 1$ \\
$D1_{(1)}$  &   $M\sim \frac{M_s}{g_s} r_1 \sim \lambda^{-\frac{2}{5}}$ & $U_{(j1)}$
                          & $S\sim \frac{m_{{\rm KK}_{(1)}}}{m_{{\rm KK}_{(j)}}}\sim  \lambda$ \\
${\rm KK}_{(1)}$  &   $M\sim \frac{M_s}{  r_1}\sim \lambda ^{-\frac{2}{5}}$ & $E\!F1_{(1j)}$ &  $S \sim \frac{m_{F1_{(1)}}}{m_{{\rm KK}_j}}\sim  \lambda$ \\
    &   & $E\!D(-1)$ &  $S \sim \frac{m_{D1_{(1)}}}{m_{F1_{(1)}}}\sim 1$ \\   
  \hline\hline
 $F1$  &   $T\sim M_s^2 \sim\lambda
         ^{-\frac{4}{5}}$ &   &  \\
  $D1$  &   $T\sim \frac{M_s^2}{ g_s}\sim\lambda
              ^{-\frac{4}{5}}$ &   &  \\
  \hline
\end{tabular}
\caption{Mass scales of light towers of particles/strings and their
  induced instanton corrections in Type IIB decompactification limit
  to 9D.}
 \label{table_emergeFb}
\end{center} 
\end{table}

This is clearly an F-theory-like  limit with the fundamental and the $D1$-string on equal footing. In contrast to the eight-dimensional limit, the $D(-1)$-instanton corrections will be fully captured by the Schwinger integral over the light states. The five towers of particle like states  $F1_{(j)}$, $D1_{(j)}$ and $D3_{(123)}$ are more massive than the species scale and are not summed over in the Schwinger integral. Moreover, the Euclidean $E\!F1$ and $E\!D1$-strings wrapping the $(23)$ plane have instanton actions $S\sim \lambda^{2}$ and are considered
as genuinely non-perturbative.

\subsubsection{Instanton expansion in the Type IIB frame}

Let us now study the induced $R^4$-term and compare it to the full result \eqref{mtheorychwingerresultstypeIIB}. The Schwinger integral for the light towers of states in the Type IIB frame takes the form
\eq{
\label{ftheoryschwinger}
a_{7}\simeq \frac{2\pi r_1}{r_2 r_3} \hat{\sum_{m_i,n_1,p_1}} \int_0^\infty  \frac{d t}{t^{\frac12}}\;
\delta({\rm BPS})\, e^{-{\pi t} \big(n_1^2 r_1^2 +p_1^2 \frac{r_1^2}{g_s^2} +\sum_i
     \frac{m_i^2}{r_i^2} \big)  }\,,
 }
where we sum over all three KK modes $m_i$ and the $F1$ and $D1$-string wrapping numbers $n_1$, $p_1$. The two simple BPS constraints read
\eq{
  m_1\, n_1 =0\,,\qquad\quad m_1\, p_1 =0\,.
}  
This integral can be evaluated following the same procedure used in the eight-dimensional case, yielding
\eq{
  \label{FtheorychwingerresultstypeIIB}
  a_7=&\frac{2 r_1^2\zeta(3)}{g_s^2}+
  \frac{8\pi r_1^2}{g_s^2} \sum_{m>0} \hat{\sum_{m_1,m_2,m_3 }} \frac{m}{l}
    K_1\big(2\pi  m l\big)\\
   &+ \frac{2 \pi^2}{3} r_1^2 +
    4\pi r_1^2 \sum_{n>0} \hat{\sum_{m_1,m_2}}
    \frac{e^{-2\pi n\, l_t }}{l_t}\\
    &+ \frac{2 \pi^2}{3}  +
    4\pi  \sum_{n>0} \hat{\sum_{m_1,m_2}}
    \frac{e^{-2\pi n\, l_u }}{l_u}\\
  &+ \frac{2\pi r_1}{  r_2 r_3}\,{\cal E}^{\rm
    SL(2)}_{\mathbf{2};s=1/2}\Big({\frac{1}{r_2},\frac{1}{r_3}}\Big)\,,
}
with the $E\!D(-1)$ and  $E\!D1_{(1j)}$-instanton actions given by \eqref{linIIB},
while those for $E\!F1_{(1j)}$ and $U_{1j}$ are given by
\eq{
  l_t= \sqrt{ m_1^2 (r_1 r_2)^2 + +m_2^2 ( r_1 r_3)^2}\,,\qquad\quad
   l_u= \sqrt{ m_1^2 \Big(\frac{r_2}{r_1}\Big)^2 + +m_2^2 \Big( \frac{r_3}{r_1}\Big)^2}\,.
}

First we observe that all the instantons from Table
\ref{table_emergeFb} are indeed present in
\eqref{FtheorychwingerresultstypeIIB}.
Comparing this to the complete $R^4$-term in Type IIB
frame \eqref{mtheorychwingerresultstypeIIB}, one realizes
that the individual contributions in \eqref{FtheorychwingerresultstypeIIB}
are  already the exact result. Hence, including the heavy towers of states that were left out has no effect on these contributions.
To see this explicitly one needs the  relation
\eq{
            \sum_{\alpha\in S}  {\cal J}^{\alpha}\Big( r_2
            r_3;\frac{r_2}{r_1}, \frac{r_3}{  r_1}, \frac{r_2 r_3}{
              g_s}\Big)\Big\vert_{{\rm no}\;(E\!D1_{(23)}, E\!F1_{(23)})}=
            4\pi  \sum_{n>0} \hat{\sum_{m_1,m_2}}
    \frac{e^{-2\pi n\, l_u }}{l_u}
}
and the expansion \cite{Obers:1999um}
\eq{
     \frac{2\pi r_1}{  r_2 r_3}\,{\cal E}^{\rm
    SL(3)}_{\mathbf{3};s=1/2}\big(r_1,{\textstyle \frac{1}{r_2},\frac{1}{r_3}}\big)=
     \frac{2 \pi^2}{3} r_1^2 +
    4\pi r_1^2 \sum_{n>0} \hat{\sum_{m_1,m_2}}
    \frac{e^{-2\pi n\, l_t }}{l_t}     + \frac{2\pi r_1}{  r_2 r_3}\,{\cal E}^{\rm
    SL(2)}_{\mathbf{2};s=1/2}\big({\textstyle \frac{1}{r_2},\frac{1}{r_3}}\big).
}
Interestingly, the F-theoretic Schwinger integral of the light
towers of states provide already all the non-instantonic
contributions
\eq{
         a^{\rm (n.I.)}_7=\frac{2 r_1^2\zeta(3)}{g_s^2}+ \frac{2 \pi^2}{3} r_1^2 + \frac{2 \pi^2}{3} \,,
}
so unlike in emergent string limits, no classical tree-level contribution is absent as in the M-theory limit. However, the full contribution is not reproduced, since the Euclidean $E\!F1_{(23)}$ and $E\!D1_{(23)}$-instantons are missing.

\subsection{Decompactification to 11D}
Like in the eight-dimensional case, there also exist more general infinite distance limits corresponding to decompactifications of $d\le 4$ dimensions. Here, we consider the decompactification of four dimensions, i.e. to M-theory in 11D. This limit is given by the scaling
\eq{
  \label{decompact11Dlimit}
       g_s\to \lambda g_s\,,\qquad r_i\to \lambda  r_i
       \,,\qquad M_s\to
       \lambda^{-{\frac{1}{5}}}  M_s
}
with $i=1,2,3$  and $\lambda\to\infty$. Then, as shown in Table \ref{table_emergecompact11D},  the set of light towers of states consists entirely of particle-like states without any accompanying string in 7D.

%%%%%%%%%%%%%%%%%%%%%%%%%
\begin{table}[ht] 
\renewcommand{\arraystretch}{1.2} 
\begin{center} 
\begin{tabular}{|l|l||l|l|} 
\hline
light towers    & mass scale & instantons & action   \\
\hline \hline
$D0$  &   $M\sim \frac{M_s}{g_s} \sim \lambda^{-\frac{6}{5}}$ & $E\!D0_{(i)}$
                          & $S\sim \frac{m_{D0}}{m_{{\rm
                            KK}_{(i)}}}\sim 1$ \\
${\rm KK}_{(i)}$  &   $M\sim \frac{M_s}{r_i}\sim \lambda ^{-\frac{6}{5}}$ &    $U_{(ij)}$ &  $S\sim  \frac{m_{{\rm KK}_{(j)}}}{m_{{\rm
                                       KK}_{(i)}}} \sim 1$ \\
  \hline
\end{tabular}
\caption{Mass scales of light towers of particles/strings and their
  induced instanton corrections for a decompactification limit to 11D.}
 \label{table_emergecompact11D}
\end{center} 
\end{table}
\noindent
Using \eqref{speciesscale} together with the M-theory relations \eqref{mparameters}, one finds for the species scale
\eq{
\Lambda_{\rm sp}\sim M_* \sim \lambda^{-\frac{8}{15}}\,,
}
which coincides with the eleven-dimensional Planck scale. It is clear that integrating out these four light towers of states one gets precisely the second line of \eqref{mtheoryschwingerresults}
\eq{
  \label{mtheoryschwingerresults2nd}
  a_7=&\frac{2\zeta(3)}{g_s^2}+
  \frac{8\pi}{g_s^2} \sum_{m>0} \hat{\sum_{m_1,m_2,m_3}} \frac{m}{l}
    K_1\big(2\pi  m l\big)
  + \frac{2\pi}{r_1r_2 r_3}{\cal E}^{\rm SL(3)}_{\mathbf{3};s=1/2}\Big(\frac{1}{r_1},\frac{1}{r_2},\frac{1}{r_3}\Big)\,,
}
giving the contribution of $E\!D0$-brane and $U_{(ij)}$-instantons. Again, this could be interpreted as a ``1-loop'' M-theory contribution, where the missing term $2\pi^2/3$ is the tree-level contribution and Euclidean $E\!M2$-branes provide the instanton corrections.

\section{Conclusions}
\label{section5}

In the present work, we have extended the analysis of the M-theoretic Emergence Proposal to general infinite distance limits across the moduli space of toroidal compactifications of M-theory. In particular, by focusing on the $1/2$-BPS protected $R^4$-term, we have been able to identify a general pattern of the non-perturbative corrections that arise at infinite distance limits by integrating out light towers of states. Our proposal can be summarized as follows: integrating out the light towers of states with mass scales not exceeding the species scale fully generates the instanton corrections whose action is bounded by \eqref{window}, i.e. the ratio of the species scale and the lightest tower in the limit. Importantly, we remark that it is the full contribution that emerges (to be compared with the contribution appearing in the full amplitude) for each instanton within the bound. This generically requires a number of non-trivial cancellations between the contributions from bound states of heavy states, both involving and not involving light states, making the contribution of the heavy states redundant. 

We have supported this statement by considering the entire eight-dimensional moduli space and a number of seven-dimensional examples. The former let us test the generality of our proposal, while the latter provided us with evidence where the cancellations happen in a more intricate manner. In every example we confirm that, indeed, the light tower of states are enough to generate the expected non-perturbative corrections, and thus, the heavy states must give mutually canceling contributions.

We have also reformulated our claim in the language of the taxonomy classification of \cite{Etheredge:2024tok}, introducing what we have coined as $S$-vectors, which encode the direction in moduli space in which the instantons become subleading. While we have not explored the geometry that may underlie these vectors, we have been able to translate the claim above in terms of the data given by the tower and species polytopes. The bound in this case is given in \eqref{taxonomywindow}. Although an explicit verification of such a claim becomes increasingly complicated in lower dimensions, we could explore any interesting phenomena it implies. For instance, it would be worthwhile to explore whether the non-string-embeddable polytopes in six dimensions of \cite{Etheredge:2024tok} lead to a contradiction with our claim, rendering them inconsistent from emergence arguments.

On the other hand, it is quite remarkable that the M-theory limit of \cite{Blumenhagen:2024ydy} remains as the sole infinite distance limit in the moduli space of toroidal compactifications of Type IIA in which the entire coupling emerges from a one-loop effect. Nonetheless, we have encountered other decompactification limits, in particular where two directions open up giving F-theory-like limits, in which no classical, tree-level term is missing. Such limits appear naturally in the classification of infinite distance limits in the vector multiplet moduli space of Calabi-Yau threefolds, although we leave their study in the context of the Emergence Proposal to possible future work.

Finally, a question that remains open is on the fate of non-BPS protected amplitudes, which in general receive contributions from every state in the theory. In the case of the M-theory limit this requires a quantum description of M-theory which is absent at the moment. For the rest of infinite distance limits, the computation of quantities not protected behind non-renormalization theorems remains a daunting task, leaving our proposal speculative. However, as a final remark, let us note that other higher-curvature corrections, which are generally less protected by supersymmetry, are given by similar modular forms (see e.g. \cite{Green:2005ba, Russo:1997mk}) and would be a natural testing ground for our claim.

\appendix
\paragraph{Acknowledgments.}
We thank Muldrow Etheredge and Alvaro Herr\'aez for enlightening discussions. We are also grateful to Antonia Paraskevopoulou for useful comments about the manuscript.

% \vspace{1cm}

% \newpage

\section{Instanton vectors in 8D}
\label{AppendixA}

In this appendix, we determine the relation between canonically and non-canonically normalized radii. In \cite{Etheredge:2024tok}, the polytopes were constructed using canonically normalized radii together with a specific choice of basis for the $\alpha$-vectors.

Following the conventions in \cite{Etheredge:2024tok}, we consider the compactification of a $D$-dimensional theory on a diagonal $k$-torus. To this end, we adopt the following metric ansatz
\begin{equation}
\label{metric}
ds_D^2 = |g_{ij}|^{-\frac{1}{d-2}}\, ds_d^2 + g_{ij}\, d\theta^i d\theta^j \,,
\qquad 
g_{ij} = \delta_{ij}\, e^{-2\rho_i}\,,
\end{equation}
where $ds_d^2$ denotes the metric of the $d$-dimensional non-compact spacetime,  while $\theta^i$ ($i=1,\dots,k$) parameterize the internal torus directions with metric $g_{ij}$.

Starting from the $D$-dimensional Einstein--Hilbert action
\begin{equation}
\label{highaction}
S_D = \frac{1}{2\kappa_D^2}\int d^D x \sqrt{-g}\, R_D\,,
\end{equation}
after dimensional reduction, this yields the effective $d$-dimensional action
\begin{equation}
\label{effectiveaction}
S_d=\frac{1}{2\kappa_d^2}\int d^dx\sqrt{-g}\left(
R - \sum_i (\nabla \rho_i)^2
-\frac{1}{d-2}\sum_{ij}\nabla \rho_i \nabla \rho_j
\right).
\end{equation}
We can directly read off the radion metric as
\begin{equation}
\label{fieldmetric}
G_{ij}=\delta_{ij}+\frac{1}{d-2}\,1_i1_j \  ,
\end{equation}
where $1_i$ denotes the vector with all components equal to one. 

The metric on the moduli space spanned by the radions $\rho_i$ is not the identity matrix. In order to obtain a canonically normalized basis, it is convenient to introduce a new set of moduli $\tilde R_i$ defined in terms of the radions $\rho_i$ as\cite{Etheredge:2024tok}
\begin{equation}
 \tilde{R}^i \equiv -\rho^i - \frac{\sum_j \rho^j}{d-2 + \sqrt{(d-2)(D-2)}} \ 1^i \,,
\end{equation}
The relation can also be inverted, allowing the original radions $\rho_i$ to be expressed in terms of the canonically normalized fields $\tilde{R}_i$. This yields
\begin{equation}
\label{rhointermsofR}
    \rho_i = - \tilde{R}_i -\frac{\sum_j\tilde{R}_j}{k}
\left( -1+ \sqrt{\frac{d-2}{D-2}}\right) 1_i \ ,
\end{equation}
where $k$ denotes the number of internal torus directions.

Now, considering the conventions used in the main text, it is
straightforward to express the masses from \eqref{massBPSstates} in
terms of the canonically normalized fields $\tilde R_i$. Let $D=11$
and a three-dimensional torus with the ansatz \eqref{metric}. This
implies that the radii in string units used throughout the text,
$r_i=M_s R_i$, is related to the new conventions through
$R_i=e^{-2\rho_i}$. Then, we can consider for example the $D0$-branes and compute their mass with respect to the canonically normalized radions to be given by
\begin{equation}
\label{MassD0}
m_{D0} = M_{\rm Pl}^{(8)}\,
e^{-\hat{\tilde R}^{11} \left(\frac{2}{3} + \frac{1}{2} \sqrt{\frac{2}{3}} \right)
- \hat{\tilde R}^1\left( -\frac{1}{3} + \frac{1}{2} \sqrt{\frac{2}{3}}\right)
- \hat{\tilde R}^2\left( -\frac{1}{3}+\frac{1}{2} \sqrt{\frac{2}{3}}\right)}\,,
\end{equation}
with $\hat{\tilde R}_i=M_{\rm Pl}^{(8)} \tilde R_i$ the canonically normalized radions in eight-dimensional Planck units.

Using the definition of $\alpha$-vector in \eqref{alphavectordefinition}, together with the mass expression 
\eqref{MassD0}, we obtain the corresponding $\alpha$-vector
\begin{equation}
\label{alphaD0}
\vec{\alpha}_{D0} =
\left(
-\frac{1}{3} + \frac{1}{2} \sqrt{\frac{2}{3}} \ ,\
-\frac{1}{3} + \frac{1}{2} \sqrt{\frac{2}{3}} \ ,\
\frac{2}{3} + \frac{1}{2} \sqrt{\frac{2}{3}}
\right).
\end{equation}
To match the ones presented in \cite{Etheredge:2024tok}, we can perform an appropriate change of basis to their conventions through the matrix transformation
\begin{align}
\label{transformationmatrix}
A = \begin{pmatrix}
\frac{1}{\sqrt{2}} & 0 & -\frac{1}{\sqrt{2}}\\
-\frac{1}{\sqrt{3}} & -\frac{1}{\sqrt{3}} & -\frac{1}{\sqrt{3}}\\
-\frac{1}{\sqrt{6}} & \frac{2}{\sqrt{6}} & -\frac{1}{\sqrt{6}}
\end{pmatrix}\,,
\end{align}
so that
\begin{equation}
    A\cdot \vec\alpha_{D0}=\left(-\frac{1}{\sqrt{2}},-\frac{1}{\sqrt{2}},-\frac{1}{\sqrt{6}}\right)\,,
\end{equation}
agreeing with conventions used throughout the text. 

Having determined the masses of the particle states in terms of canonical normalized radions, we can also derive the actions of the corresponding instantons and define their associated $\alpha$-vectors, as introduced in \eqref{instantonvector}. 
As an example, consider the complex structure instanton $U_{(12)}$, whose action is given by
\begin{equation}
S_{U_{21}}
= \frac{\tilde{R}_2}{\tilde{R}_1}= e^{R^2 - R^1}\Rightarrow \vec S_{U_{21}}=(1,-1,0)\,.
\end{equation}
To express it in the appropriate basis, we multiply by \eqref{transformationmatrix} and obtain
\begin{equation}
\vec S_{U_{12}} =
\left(\frac{1}{\sqrt{2}},\,0,\,-\sqrt{\frac{3}{2}}\right)\,.
\end{equation}
Repeating the same procedure for the remaining instantons yields 
\begin{equation}
    \vec{S}_{EF1_{12}}=\left(0,{\sqrt{2}},0\right)\,,\quad
    \vec{S}_{ED0_1}=\left(-{\sqrt{2}},0,0\right)\,,\quad\vec{S}_{ED0_2}=\left(-\frac{1}{\sqrt{2}},0,-\sqrt{\frac{3}{2}}\right)\,.
\end{equation}

\newpage

\bibliography{references} 
\bibliographystyle{utphys}

\end{document}